\documentclass[letter, a4paper, twocolumn]{aa}
\makeatletter
\renewcommand*\aa@pageof{, page \thepage{} of \pageref*{LastPage}}
\makeatother

\pdfoutput=1
\usepackage{xcolor, graphicx, amssymb, mathrsfs, placeins, xspace}

\usepackage{natbib}
\usepackage{txfonts}
\usepackage{comment}
\usepackage[colorlinks=true, allcolors=blue]{hyperref}
\usepackage{orcidlink}

\newcommand{\percent}{\ensuremath{\%}}
\newcommand{\Msun}{M_\odot}

\newcommand{\Mgas}{M_\mathrm{gas}}
\newcommand{\Mstar}{M_\mathrm{star}}

\newcommand{\Mbar}{M_\mathrm{bar}}
\newcommand{\gbar}{g_\mathrm{bar}}

\newcommand{\gbarmed}{\langle\gbar\rangle}

\newcommand{\Reff}{R_{\mathrm{e}}}
\newcommand{\kms}{\mathrm{km~s^{-1}}}
\newcommand{\aZero}{a_0}
\newcommand{\Sersic}{S\'ersic}

\begin{document}

%%%%%%%%%%%%%%%%%%%%%%%%%%%%%%%%%%%%%%%%%%%%%%%%%%%%%%%%%
%                 Title & Authors
%%%%%%%%%%%%%%%%%%%%%%%%%%%%%%%%%%%%%%%%%%%%%%%%%%%%%%%%%
\title{The Baryonic Faber-Jackson Relation and Fundamental Plane of Galaxy Groups, Elliptical Galaxies, and Dwarf Galaxies}

   \date{Received 18 February 2026; Accepted 11 May 2026}

\titlerunning{BFJR}
\authorrunning{Yong Tian et al.}

\author{Yong Tian\,\orcidlink{0000-0001-9962-1816}\inst{1,2}\fnmsep\thanks{yongtian@phy.ncu.edu.tw} 
    \and
        Federico Lelli\,\orcidlink{0000-0002-9024-9883}\inst{3}\thanks{federico.lelli@inaf.it}
    \and
        Marcel S. Pawlowski\,\orcidlink{0000-0002-9197-9300}\inst{4}
    \and
        Stacy McGaugh\,\orcidlink{0000-0002-9762-0980}\inst{5}
    \and
        Yi Duann\,\orcidlink{0000-0002-4260-385X}\inst{6}
    \and
        Kyu-Hyun Chae\,\orcidlink{0000-0002-6016-2736}\inst{2}
    \and
        Enrico Di Teodoro\,\orcidlink{0000-0003-4019-0673}\inst{7,3}
    \and
        Konstantin Haubner\,\orcidlink{0009-0007-7808-4653}\inst{3,7}
    \and
        Meng Hua Kuo\,\inst{8, 9}
    \and
        Chung-Ming Ko\,\orcidlink{0000-0002-6459-4763}\inst{10,11}
        }

\institute{Department of Physics, National Central University, Taoyuan 320317, Taiwan
        \and
            Department of Physics and Astronomy, Sejong University, 209 Neungdong-ro Gwangjin-gu, Seoul 05006, Republic of Korea
        \and
            INAF – Arcetri Astrophysical Observatory, Viale Enrico Fermi 5, 50125 Florence, Italy
        \and
            Leibniz-Institute for Astrophysics, An der Sternwarte 16, 14482 Potsdam, Germany
        \and
            Department of Astronomy, Case Western Reserve University, 10900 Euclid Avenue, Cleveland, OH 44106, USA
        \and
            Globe Institute--Center for Star and Planet Formation, University of Copenhagen, Øster Voldgade 5-7, Copenhagen 1350, Denmark 
        \and
            Dipartimento di Fisica e Astronomia, Università degli Studi di Firenze, Via G. Sansone 1, 50019 Sesto Fiorentino, Firenze, Italy
        \and
            Department of Economics, National Central University, Taoyuan 320317, Taiwan
        \and
            Department of Biomedical Sciences and Engineering, National Central University, Taoyuan 320317, Taiwan
        \and
            Institute of Astronomy, National Central University, Taoyuan 320317, Taiwan
        \and
            Department of Physics and Center for Complex Systems, National Central University, Taoyuan 320317, Taiwan
            }

%%%%%%%%%%%%%%%%%%%%%%%%%%%%%%%%%%%%%%%%%%%%%%%%%%%%%%%%%
%
%                 Abstract & Keywords
%
%%%%%%%%%%%%%%%%%%%%%%%%%%%%%%%%%%%%%%%%%%%%%%%%%%%%%%%%%

\abstract{
The baryonic Faber-Jackson relation (BFJR) links the baryonic mass of pressure-supported systems to their mean velocity dispersion. For elliptical galaxies, the BFJR is thought to be a projection of the fundamental plane (FP), which includes the stellar half-mass radius as a third variable. We study the BFJR and FP across eight orders of magnitude in baryonic mass, encompassing galaxy groups, ellipticals, dwarf ellipticals, and dwarf spheroidals. We compile and homogenize data for 1400 pressure-supported systems and measure their mean internal baryonic acceleration $\langle g_\mathrm{bar}\rangle$. We find that the properties of the BFJR and FP systematically depend on the internal acceleration of the sampled systems, with a transition around the acceleration scale $a_0\simeq 1.2\times10^{-10}$ m s$^{-2}$. For low-acceleration systems with $\langle g_\mathrm{bar}\rangle < 0.6\,a_0$ (dwarf galaxies and galaxy groups), the BFJR relation takes the form $\log_{10}(M_\mathrm{bar}/M_\odot) = (4.19 \pm 0.10) \log_{10}(\sigma_{\rm los}/\kms) + (2.55^{+0.16}_{-0.16})$. The FP expected from the Newtonian virial theorem is followed by high-acceleration systems (massive ellipticals with $\langle g_\mathrm{bar}\rangle \gtrsim 6 \,a_0$), whereas low-acceleration systems deviate from the FP at both low masses (dwarf galaxies) and high masses (galaxy groups). Our results generally agree with the expectations of modified Newtonian dynamics (MOND): high-acceleration systems follow the Newtonian virial theorem in which a radial variable explicitly appears (the FP), while low-acceleration systems follow the MOND virial theorem in which the radial dependence disappears (the BFJR). On average, the MOND external field effect seems to play a secondary role in dwarf galaxies in galaxy groups and clusters.
}
\keywords{galaxies: kinematics and dynamics – galaxies: elliptical and lenticular, cD – galaxies: dwarf – dark matter – gravitation}

\maketitle
\nolinenumbers

%%%%%%%%%%%%%%%%%%%%%%%%%%%%%%%%%%%%%%%%%%%%%%%%%%%%%%%%%
%
%                Section: Introduction
%
%%%%%%%%%%%%%%%%%%%%%%%%%%%%%%%%%%%%%%%%%%%%%%%%%%%%%%%%%
\section{Introduction}\label{sec:intro}

Scaling relations between the visible mass of galaxies and their observed kinematics provide fundamental insights into the interplay between baryons, dark matter (DM), and standard gravity. A classic example is the Faber-Jackson Relation (FJR), which describes a correlation between the luminosity and stellar velocity dispersion of elliptical galaxies \citep{FJ1976}. An extension of the classic FJR is the baryonic Faber-Jackson relation (BFJR), which links the total baryonic mass $\Mbar$ (stars plus gas) to the line-of-sight velocity dispersion $\sigma_{\rm los}$ \citep[e.g.,][]{Sanders2010, FM2012}. This is analogous to superseding the classic Tully-Fisher relation with the baryonic Tully-Fisher relation \citep[BTFR;][]{McGaugh2000, Lelli2016, Lelli2019}. For elliptical galaxies, the FJR is generally thought to be a projection of the so-called Fundamental Plane \citep[FP;][]{Djorgovski1987, Dressler1987}, which adds the stellar effective radius ($R_{\rm e}$) as a third variable. The FP is expected from the Newtonian virial theorem ($M\propto R \cdot \sigma_{\rm los}^2$) but its observed parameters show a ``tilt'' with respect to the virial parameters \citep[e.g.,][]{Ciotti1996}, which may be driven by variations in the stellar mass-to-light ratio, inner DM fractions, orbital anisotropy, or structural non-homology \citep[e.g.,][]{Cappellari2006, Bolton2007, Cappellari2013b}.

In the $\Lambda$ Cold Dark Matter ($\Lambda$CDM) cosmological paradigm, scaling relations must emerge from a combination of complex stochastic physical processes, including the hierarchical growth of DM halos, accretion and cooling of gas, star formation, feedback from supernovae and active galactic nuclei, and the feedback effects on the structural properties of DM halos \citep{Pillepich2018}. Within the DM framework, this leads to a ``fine-tuning'' problem: a precise coupling between baryons and DM is required to place DM-dominated dwarf galaxies and baryon-dominated giant galaxies on the same relations~\citep[e.g.,][]{FM2012, Desmond2017, Lelli2022}.

The main alternative to particle dark matter is MOND~\citep[MOdified Newtonian Dynamics or MilgrOmiaN Dynamics,][]{Milgrom1983}. 
MOND postulates that the non-relativistic laws of dynamics (gravity or inertia) are modified at accelerations below a characteristic scale $\aZero = 1.2 \times 10^{-10}\mathrm{m\,s}^{-2}$ \citep[see][for reviews]{FM2012, Banik2022}. In particular, the MOND virial theorem \citep{Milgrom1984, Milgrom2014a} implies a universal scaling for isolated, self-gravitating, virialized systems in the deep-MOND regime (internal accelerations $g \ll \aZero$):
\begin{equation}
\centering
\Mbar \propto \sigma_{\rm 3D}^4/(G\aZero),
\label{eq:MOND_BFJR}
\end{equation}
where $\sigma_{\rm 3D}$ is the 3D mass-weighted velocity dispersion of the system. Eq.\,(\ref{eq:MOND_BFJR}) is markedly different from the Newtonian virial theorem because it does not contain any dependence on the characteristic size of the system. It is expected to hold only for systems in which $g \ll \aZero$, while systems in which $g \gg \aZero$ should follow the Newtonian relation with no DM. The general MOND paradigm predicts the slope of the BFJR to be exactly four. The value of the intercept depends on the specific MOND theory, but differences are of order $O(1)$~\citep{Milgrom2014b, Milgrom2025}.

To study the properties of the BFJR and FP, we compiled an unprecedented dataset spanning eight orders of magnitude in baryonic mass, from dwarf galaxies to massive ellipticals and galaxy groups. 
We examined how the parameters of the BFJR and FP vary with internal acceleration, providing a stringent test for both $\Lambda$CDM theories of galaxy formation and MOND.

%%%%%%%%%%%%%%%%%%%%%%%%%%%%%%%%%%%%%%%%%%%%%%%%%%%%%%%%%
%
%                Section: Data & Analysis
%
%%%%%%%%%%%%%%%%%%%%%%%%%%%%%%%%%%%%%%%%%%%%%%%%%%%%%%%%%
\section{Data analysis}\label{sec:data_methods}

We built a comprehensive sample covering different types of pressure-supported systems, including (1) 63 galaxy groups in the Local Supercluster~\citep{MK2011, Milgrom2019, ST2024}; (2) 1218 elliptical galaxies from the Mapping Nearby Galaxies at APO survey~\citep[MaNGA;e.g.,][]{Bundy2015, Duann2023}; (3) 26 elliptical galaxies from the ATLAS$^{3\rm D}$ survey~\citep{Cappellari2011};  (4) 34 dwarf ellipticals in the Virgo cluster~\citep{Toloba2014}; (5) 31 dwarf ellipticals in the Fornax cluster~\citep{Eftekhari2022}; and (6) 28 dwarf spheroidals in the Local Group~\citep{Lelli2017}. This combined sample spans the ranges
$\Mbar \simeq 10^5 - 10^{13}\,\Msun$ and $\sigma_{\rm los} \simeq 10 - 300\,\kms$. 
A detailed description of each sub-sample and processing methods is provided in Appendix \ref{app:samples}.

We homogenized the definitions of key quantities across all datasets. The total baryonic mass is $\Mbar = \Mstar + \Mgas$. For dwarf galaxies with negligible gas content, $\Mbar \simeq \Mstar$ (Kroupa IMF). For ellipticals and galaxy groups, we include the hot gas mass with a median gas-to-baryon mass fraction of about $8\%$~\citep{ST2024}. The line-of-sight velocity dispersion $\sigma_{\rm los}$ is taken as the stellar velocity dispersion measured within one effective radius ($\sigma_{\rm e}$) for individual galaxies, and as the velocity dispersion of member galaxies with available redshifts
%(within the mean radius) 
for galaxy groups, using the robust biweight estimator. The same member galaxies are also used to estimate the mean projected radius of the group.

For each system, we compute the Newtonian gravitational acceleration due to the observed distribution of baryons, $\gbar(r)$, assuming spherical symmetry. For elliptical and dwarf galaxies, the radial variation of $\gbar$ is computed by deprojecting a \Sersic\ profile, using the observed \Sersic\ index, effective radius, and stellar mass~\citep{Dominguez2022, Krajnovic2013, Toloba2014, Eftekhari2022}. Then, we consider the median baryonic acceleration within one effective radius $\gbarmed$ as the characteristic internal acceleration of each system. Comparisons with non-spherical estimates for available datasets suggest that deviations remain within $\sim$30\percent, which we consider acceptable for this homogeneous analysis. For galaxy groups, the median acceleration is estimated within the mean projected radius, using the spatially resolved distribution of member galaxies. This characteristic acceleration, $\gbarmed$, serves as the principal parameter for selecting low-acceleration sub-samples throughout our study.

We model the BFJR in logarithm space as a linear relation: $ y = m\,x + b$ with $y=\log_{10}(\Mbar/\Msun)$ and $x=\log_{10}(\sigma_{\rm e}/\kms)$. We fit the data using the BayesLineFit software~\citep{Lelli2019}, which implements a Markov Chain Monte Carlo (MCMC) method considering errors on both variables as well as intrinsic scatter ($\sigma_\mathrm{int}$). The intrinsic scatter is assumed to be Gaussian and can be defined either in the vertical direction (along the y variable) or in the orthogonal one (perpendicular to the best-fit line); we explore both options.

%%%%%%%%%%%%%%%%%%%%%%%%%%%%%%%%%%%%%%%%%%%%%%%%%%%%%%%%%
%
%                Section: Results
%
%%%%%%%%%%%%%%%%%%%%%%%%%%%%%%%%%%%%%%%%%%%%%%%%%%%%%%%%%
\begin{figure*}[!htb]
\centering
\includegraphics[width=0.59\columnwidth]{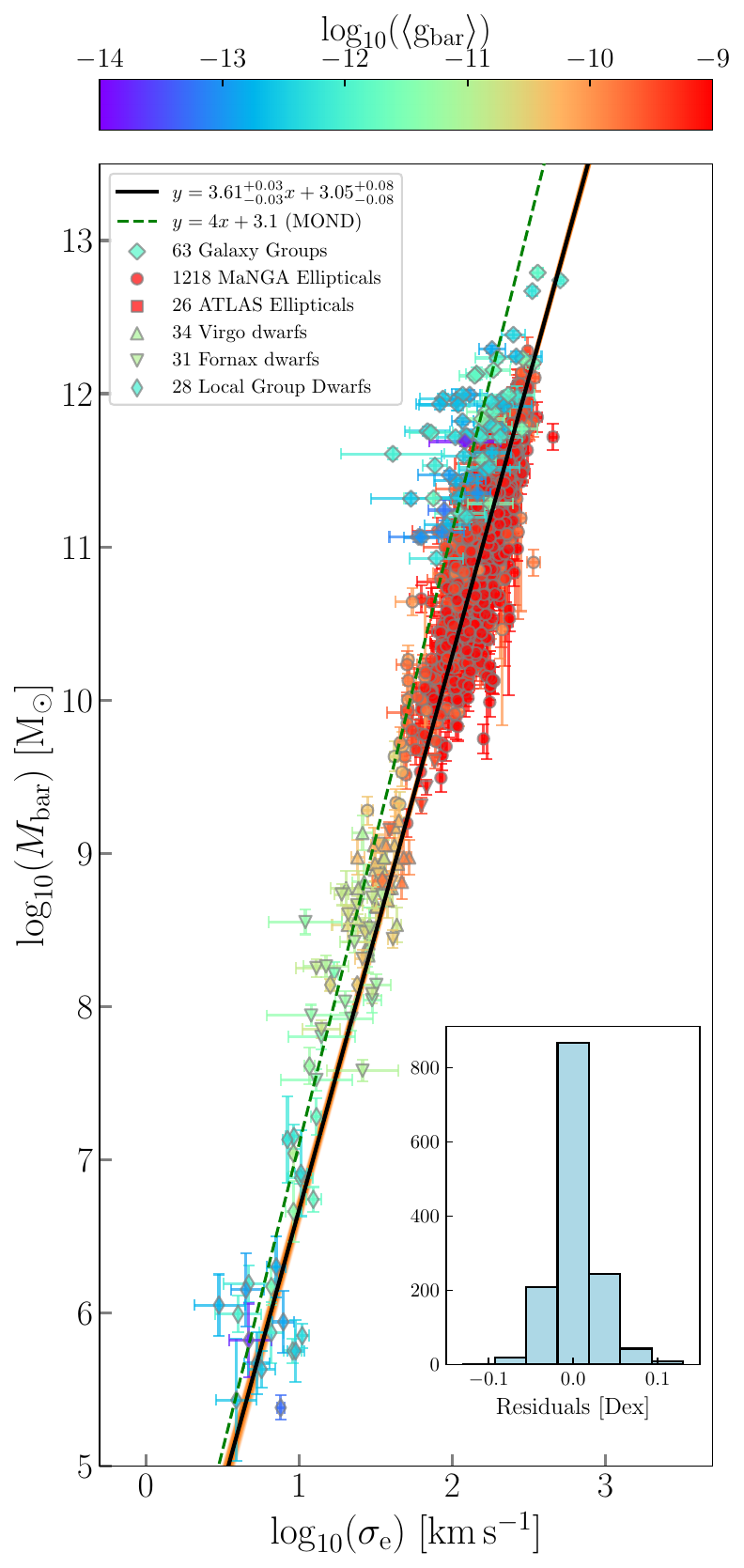}
\includegraphics[width=0.59\columnwidth]{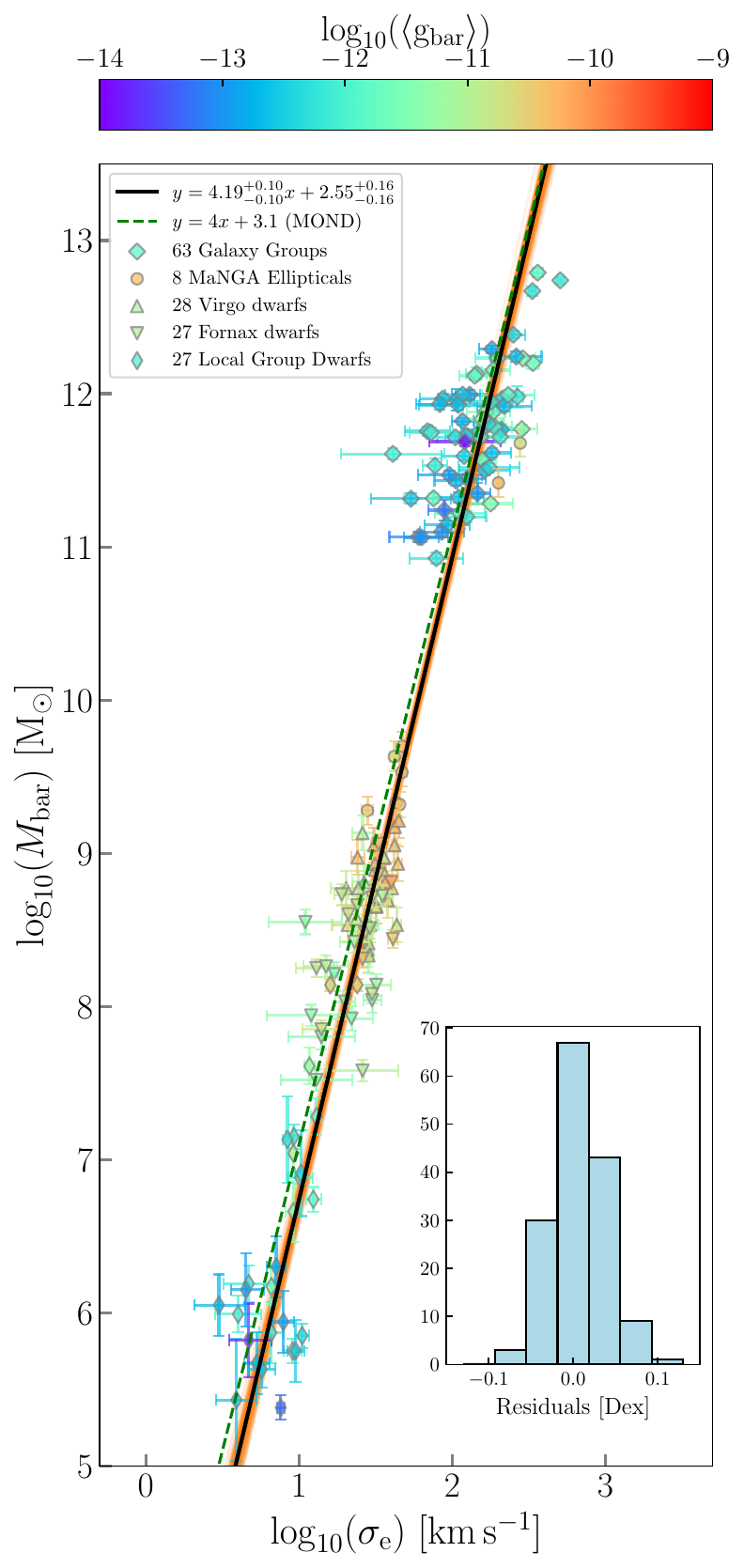}
\includegraphics[width=0.59\columnwidth]{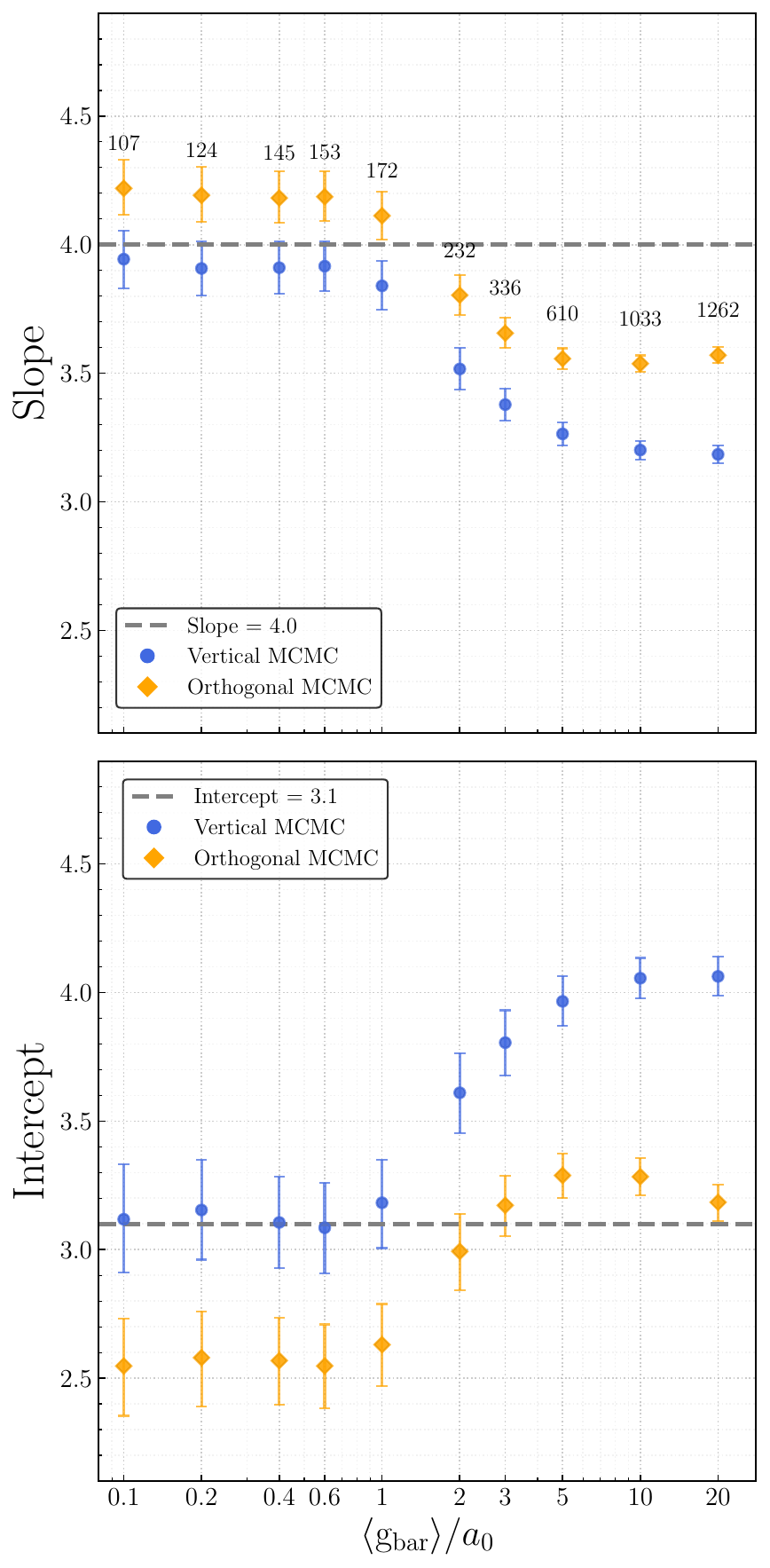}
\caption{The BFJR in galaxy groups, elliptical galaxies, and dwarf galaxies.
\textit{Left panel:} Total baryonic mass $\Mbar$ versus velocity dispersion within the effective radius $\sigma_{\rm e}$ for the full sample. 
The data points are color-coded by the internal median baryonic acceleration $\gbarmed$. 
\textit{Middle panel:} The BFJR for the low-acceleration sub-sample ($\gbarmed < 0.6\aZero$) only. In both panels, the green dashed line shows the MOND prediction in the low-acceleration regime, the black solid line is the best fit from the orthogonal MCMC, and the orange region is its 1$\sigma$ credible interval.
\textit{Right panel:} Variation of the fitted parameters with the acceleration cut-off value $\gbarmed/\aZero$:
slope $m$ (top) and intercept $b$ (bottom). Orange diamonds are the result from orthogonal MCMC fitting, blue circles from vertical MCMC fitting. The number of objects at each cut-off is listed in the right-upper panel. The horizontal dashed lines mark the theoretical expectations from MOND modified gravity theories (slope=4, intercept=3.1).}
\label{fig:BFJR}
\end{figure*}

\section{Results: Acceleration-dependent relations}\label{sec:results}

\subsection{Baryonic Faber-Jackson relation}
The BFJR is examined for the full sample in Figure~\ref{fig:BFJR} (left panel), color-coding the points by internal acceleration $\gbarmed$. All data points form a broad correlation across eight orders of magnitude in mass, but high-acceleration systems are clearly shifted towards higher $\sigma_{\rm e}$ for a given baryonic mass.

This visual trend suggests that the properties of the BFJR are not universal but instead depend on $\gbarmed$. To test this hypothesis, we fitted various sub-samples for which $\gbarmed < X\,a_0$, where $X$ is a number ranging from 0.1 to 20. Figure~\ref{fig:BFJR} (right panel) illustrates how the fitted parameters (slope $m$ and intercept $b$) vary as a function of the $\gbarmed/\aZero$ threshold. We do not show the intrinsic scatter because the various sub-samples have different sizes and heterogeneous error estimates, so comparing the intrinsic scatters can be misleading. As we restrict the sample to progressively lower accelerations, the slope ($m$) steadily converges toward $\sim$4 and the intercept ($b$) becomes stable. This convergence suggests that low-acceleration and high-acceleration systems follow a different BFJR. This is confirmed in Appendix\,\ref{app:High_acc}, in which we perform the same exercise for sub-samples for which $\gbarmed>X a_0$.

We select $\gbarmed < 0.6\,\aZero$  as a practical threshold to compromise between sample statistics (153 systems) and convergence of the best-fit results. The resulting BFJR is shown in Figure 1 (middle panel).The best-fit relation is
\begin{equation}
 \log_{10}\left(\frac{\Mbar}{\Msun}\right) = (4.19 \pm 0.10) \log_{10}\left(\frac{\sigma_{\rm e}}{\kms}\right) + (2.55^{+0.16}_{-0.16})\,.
 \label{eq:bestfit}
\end{equation}
The corner plots of both vertical and orthogonal MCMC analyses are shown in Appendix~\ref{app:corner}. We explore possible residual correlations in Appendix~\ref{app:residuals}. Importantly, the residuals show no correlation with effective radius, indicating that it is not possible to decrease the observed scatter with a third structural variable, contrary to the case of the BFJR of high-acceleration systems (see Fig.\,\ref{fig:Res_high}), see the Sect.\,\ref{sec:discussion} for further discussion.

\subsection{Fundamental plane}\label{app:FP}

\begin{figure}[!htb]
  \centering
  \includegraphics[width=0.85\columnwidth]{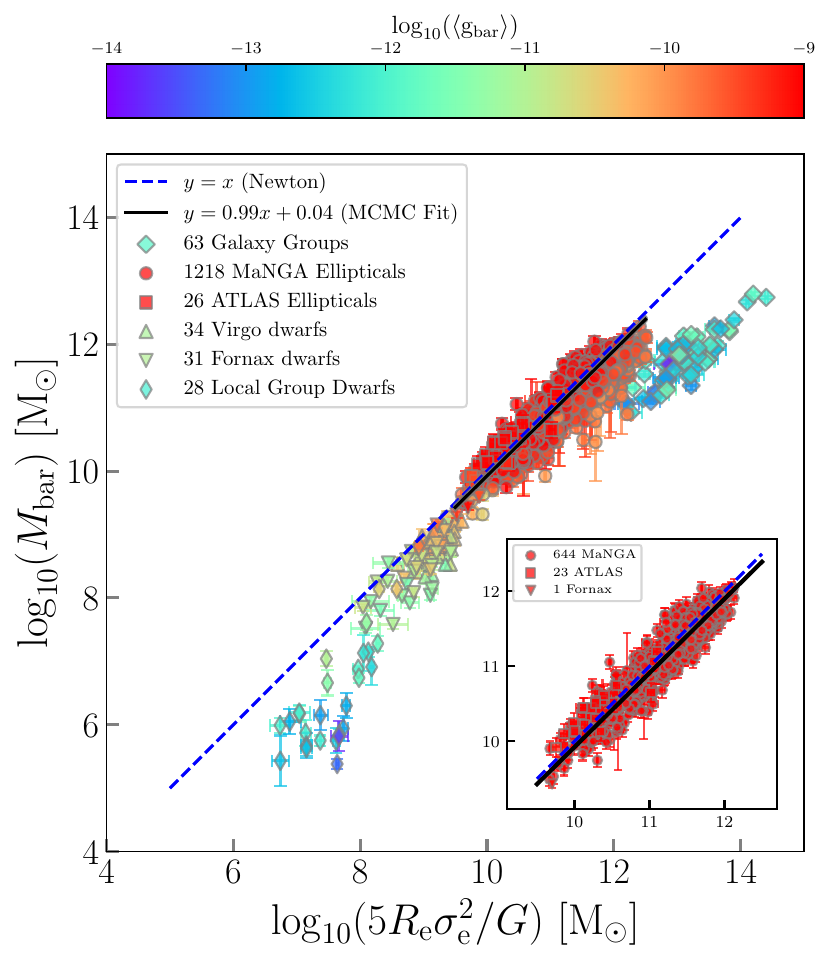}
  \caption{The FP for pressure-supported systems, including galaxy groups, ellipticals, and dwarf galaxies. The $x$-axis shows the expected Newtonian dynamical mass $\log_{10}(5 \Reff \sigma_{\rm e}^2/G)$, while the $y$-axis gives the observed baryonic mass $\log_{10}(\Mbar)$. The symbols are color-coded by the median baryonic acceleration within the effective radius. The Newtonian expectations (dashed line) are followed only by high-acceleration systems with $\gbarmed > \aZero$, while low-acceleration systems systematically depart from it. The inset panel presents the MCMC analysis for the subsample restricted to systems with $\gbarmed>6\aZero$.}
  \label{fig:FP}
\end{figure}

Figure~\ref{fig:FP} shows the baryonic FP considering $\Mbar$ versus the Newtonian virial estimator $5\Reff\sigma^2_{\mathrm{e}}/G$. The factor of 5 is adopted following~\citet{Cappellari2013a}. If no DM is present, the data should lie on the line of unity according to the Newtonian virial theorem. Massive ellipticals with high $\gbarmed$ follow the Newtonian expectation, while systems with low $\gbarmed$ systematically deviate at both low masses (dwarf galaxies) and high masses (galaxy groups). Performing a Bayesian orthogonal regression on the subsample with $\gbarmed > 6\,\aZero$, we find
\begin{equation}
    \log_{10}\left(\frac{\Mbar}{\Msun}\right) = (0.99 \pm 0.01) \log_{10}\left(\frac{5\Reff\sigma^2_{\mathrm{e}}}{G\Msun}\right) + (0.04 \pm 0.15)\,.
\end{equation}

%%%%%%%%%%%%%%%%%%%%%%%%%%%%%%%%%%%%%%%%%%%%%%%%%%%%%%%%%
%
%                Section: Discussion
%
%%%%%%%%%%%%%%%%%%%%%%%%%%%%%%%%%%%%%%%%%%%%%%%%%%%%%%%%%
\section{Discussion}\label{sec:discussion}

\subsection{Consistency with the MOND paradigm}
Our studies empirically corroborate three predictions of MOND:
\begin{enumerate}

\item \textbf{Acceleration dependence:} High-acceleration systems (elliptical galaxies) follow the Newtonian FP with no need of DM, while low-acceleration systems (dwarf galaxies and galaxy groups) deviate from it and require large amounts of DM. Yet, low-acceleration systems define a linear BFRJ with small scatter. This confirms the MOND prediction that $\aZero$ marks a transition scale below which the dynamical behavior of galaxies and galaxy systems fundamentally changes.

\item \textbf{Slope:} Our measured slopes of $4.19 \pm 0.10$ (orthogonal fit) and $3.92 \pm 0.10$ (vertical fit) are consistent with the MOND prediction of $m=4$ within the 95\percent ($\sim 2\sigma$) and 68\percent ($\sim 1\sigma$) confidence intervals, respectively.

\item \textbf{Normalization}: 
In the specific cases of the non-relativistic modified gravity theories AQUAL~\citep{BM1984} and QUMOND~\citep{Milgrom2010}, the proportionality factor in Eq.\,(\ref{eq:MOND_BFJR}) is exactly 9/4. Assuming orbital isotropy, we have $\sigma_{\rm 3D} = \sqrt{3}\sigma_{\rm los}$ and the proportionality factor becomes 81/4. The best-fit intercept of the low-acceleration BFJR is statistically consistent with this predicted value.
\end{enumerate}

The dwarf galaxies in our sample are either satellites of massive spirals (the Milky Way and Andromeda) or in galaxy clusters (Virgo and Fornax), so they may be affected by the MOND external field effect \citep[EFE,][]{BM1984}. The Newtonian external field experienced by these dwarfs is around $\sim 0.01 - 0.1\aZero$, less than our acceleration cut ($\sim 0.6\aZero$). Similarly, galaxy groups may be affected by the EFE due to the large-scale structure of the Universe, which becomes relevant for $\gbarmed\lesssim0.01\aZero$ \citep{Chae2021, Kelleher2024}. The fact that the low-acceleration BFJR agrees with the MOND prediction for isolated systems suggests that the EFE must play a secondary subtle role. Interestingly, the residuals around the BFJR show a weak correlation with $\gbarmed$ (see Fig.\,\ref{fig:Res}). This is qualitatively consistent with the EFE because the systems with the lowest internal accelerations could be more Newtonian than MONDian, so display a lower $\sigma_{\rm e}$ at fixed $M_{\rm bar}$ (or higher $M_{\rm bar}$ at fixed $\sigma_{\rm e}$). Ideally, one would like to plot the BFJR residuals against $\gbarmed/g_{\rm ext}$ where $g_{\rm ext}$ is the baryonic external field felt by each system. This requires a more accurate study of the environment of each dwarf galaxy and each galaxy group.

\subsection{Implications for galaxy formation models}

In a $\Lambda$CDM context, it is surprising that dwarf galaxies and galaxy groups lie on the same BFJR despite being totally different systems that are shaped by very different physical processes. The small scatter ($\sim0.11$ dex) observed across $\sim 8$ orders of magnitude in mass leaves little room for stochastic variation in these processes~\citep{Desmond2017}. Even if we focus only on galaxies, the tightness of the BFJR demands finely tuned feedback processes (e.g., from supernovae and AGN) to regulate star formation and set a precise DM fraction as a function of mass. Future work may compare our results to state-of-the-art cosmological simulations, such as EAGLE~\citep{Crain2015}, BAHAMAS~\citep{McCarthy2017}, and IllustrisTNG~\citep{Nelson2019}. In general, it is unclear why the formation and evolution of galaxy groups and dwarf galaxies in $\Lambda$CDM should conspire to resemble the a-priori prediction of MOND.

%%%%%%%%%%%%%%%%%%%%%%%%%%%%%%%%%%%%%%%%%%%%%%%%%%%%%%%%%
%
%                Section: Conclusions
%
%%%%%%%%%%%%%%%%%%%%%%%%%%%%%%%%%%%%%%%%%%%%%%%%%%%%%%%%%
\section{Conclusions}

By compiling a comprehensive sample of pressure-supported systems spanning about eight orders of magnitude in mass, we find that the properties of the BFJR and of the FP systematically change with the mean internal acceleration $\gbarmed$. In the low-acceleration regime ($\gbarmed < 0.6 \aZero$), the BFJR converges to a tight power law, $\Mbar \propto \sigma_{\rm e}^{4.19\pm0.10}$. In the high-acceleration regime ($\gbarmed > 6 \aZero$), the FP offers a superior description of the data compared to the BFJR. These findings are in excellent agreement with the predictions of MOND, with the characteristic acceleration scale $\aZero$ naturally accounting for the observed behavior.

Overall, the BFJR and FP emerge as fundamental scaling relations of galaxies and galaxy groups. Their tightness and acceleration dependence pose a significant fine-tuning challenge for models within the standard cosmological framework, and provide a powerful empirical testbed for distinguishing between competing theories of gravity and galaxy formation.

%%%%%%%%%%%%%%%%%%%%%%%%%%%%%%%%%%%%%%%%%%%%%%%%%%%%%%%%%
%
%                Acknowledgments
%
%%%%%%%%%%%%%%%%%%%%%%%%%%%%%%%%%%%%%%%%%%%%%%%%%%%%%%%%%
\begin{acknowledgements}
We thank the referee for the feedback and suggestions.
We also thank Moti Milgrom and Pradyumna Sadhu. 
YT acknowledges the Taiwan National Science and Technology Council (NSTC) grants 110-2112-M-008-015-MY3 and 114-2112-M-008-024-MY3.
YT and KHC acknowledge the National Research Foundation of Korea (grant no. NRF-2022R1A2C1092306).
MSP acknowledges funding via a Leibniz-Junior Research Group (project number J94/2020).
SSM is supported in part by NASA ADAP grant 80NSSC19k0570 and also acknowledges support from NSF PHY-1911909.
YD is supported by the Postdoctoral Research Abroad Program (PRAP) grant NSTC 114-2917-I-564-044 and NSTC 114-2124-M-008-003.
EDT is supported by the European Research Council (ERC) under grant agreement No. 101040751.
MHK was supported by NSTC grant 110-2112-M-008-015-MY3.
CMK is supported by the Taiwan NSTC 114-2112-M-008-018.
\end{acknowledgements}

%%%%%%%%%%%%%%%%%%%%%%%%%%%%%%%%%%%%%%%%%%%%%%%%%%%%%%%%%
%
%                Bibliography
%
%%%%%%%%%%%%%%%%%%%%%%%%%%%%%%%%%%%%%%%%%%%%%%%%%%%%%%%%%
\bibliographystyle{aa}
\bibliography{BFJR}

@ARTICLE{McGaugh26,
       author = {{McGaugh}, Stacy S. and {Mistele}, Tobias and {Duey}, Francis and {Haubner}, Konstantin and {Lelli}, Federico and {Schombert}, James M. and {Li}, Pengfei},
        title = "{The Baryonic Mass─Halo Mass Relation of Extragalactic Systems}",
      journal = {\apj},
         year = 2026,
        month = apr,
       volume = {1001},
       number = {1},
          eid = {65},
        pages = {65},
          doi = {10.3847/1538-4357/ae4ecc},
archivePrefix = {arXiv},
       eprint = {2603.06479},
 primaryClass = {astro-ph.GA},
       adsurl = {https://ui.adsabs.harvard.edu/abs/2026ApJ..1001...65M}
}

@ARTICLE{Banik2022,
       author = {{Banik}, Indranil and {Zhao}, Hongsheng},
        title = "{From Galactic Bars to the Hubble Tension: Weighing Up the Astrophysical Evidence for Milgromian Gravity}",
      journal = {Symmetry},
         year = 2022,
        month = jun,
       volume = {14},
       number = {7},
          eid = {1331},
        pages = {1331},
          doi = {10.3390/sym14071331},
archivePrefix = {arXiv},
       eprint = {2110.06936},
 primaryClass = {astro-ph.CO},
       adsurl = {https://ui.adsabs.harvard.edu/abs/2022Symm...14.1331B}
}

@ARTICLE{BM1984,
       author = {{Bekenstein}, J. and {Milgrom}, M.},
        title = "{Does the missing mass problem signal the breakdown of Newtonian gravity?}",
      journal = {\apj},
         year = 1984,
        month = nov,
       volume = {286},
        pages = {7-14},
          doi = {10.1086/162570},
       adsurl = {https://ui.adsabs.harvard.edu/abs/1984ApJ...286....7B}
}

@ARTICLE{Bundy2015,
       author = {{Bundy}, K. and {Bershady}, M.~A. and {Law}, D.~R. and {Yan}, R. and {Drory}, N. and {MacDonald}, N. and {Wake}, D.~A. and {Cherinka}, B. and {S{\'a}nchez-Gallego}, J.~R. and {Weijmans}, A.-M. and {Thomas}, D. and {Tremonti}, C. and {Masters}, K. and {Coccato}, L. and {Diamond-Stanic}, A.~M. and {Arag{\'o}n-Salamanca}, A. and {Avila-Reese}, V. and {Badenes}, C. and {Falc{\'o}n-Barroso}, J. and {Belfiore}, F. and {Bizyaev}, D. and {Blanc}, G.~A. and {Bland-Hawthorn}, J. and {Blanton}, M.~R. and {Brownstein}, J.~R. and {Byler}, N. and {Cappellari}, M. and {Conroy}, C. and {Dutton}, A.~A. and {Emsellem}, E. and {Etherington}, J. and {Frinchaboy}, P.~M. and {Fu}, H. and {Goddard}, D. and {Goldstein}, J. and {Graves}, G.~J. and {Gunn}, J.~E. and {Harding}, P. and {Jones}, A. and {Heckman}, T. and {Hogg}, D.~W. and {Hutchinson}, T. and {Jaehnig}, K. and {Johnson}, J.~A. and {Kauffmann}, G. and {Kinemuchi}, K. and {Klaene}, M. and {Knapen}, J.~H. and {Leauthaud}, A. and {Li}, C. and {Lin}, L. and {Maiolino}, R. and {Malanushenko}, V. and {Malanushenko}, E. and {Mao}, S. and {Maraston}, C. and {McDermid}, R.~M. and {Merrifield}, M.~R. and {Nichol}, R.~C. and {Oravetz}, D. and {Pan}, K. and {Parejko}, J.~K. and {Pradhan}, B. and {Raddick}, M.~J. and {Rix}, H.-W. and {Riffel}, R. and {Riffel}, R.~A. and {Roman-Lopes}, A. and {Rom{\'a}n-Z{\'u}{\~n}iga}, C. and {Rosado}, M. and {Rubin}, K.~H.~R. and {S{\'a}nchez}, S.~F. and {Schlegel}, D.~J. and {Simmons}, A. and {Simonian}, G. and {Thanjavur}, K. and {Tinker}, J.~L. and {van den Bosch}, R.~C.~E. and {Westfall}, K.~B. and {Wilkinson}, D. and {Willis}, D. and {Wilson}, J. and {Zhang}, K.},
        title = "{Overview of the SDSS-IV MaNGA Survey: Mapping nearby Galaxies at Apache Point Observatory}",
      journal = {\apj},
         year = 2015,
        month = jan,
       volume = {798},
       number = {1},
          eid = {7},
        pages = {7},
          doi = {10.1088/0004-637X/798/1/7},
archivePrefix = {arXiv},
       eprint = {1412.1482},
 primaryClass = {astro-ph.GA},
       adsurl = {https://ui.adsabs.harvard.edu/abs/2015ApJ...798....7B}
}

@ARTICLE{Kelleher2024,
       author = {{Kelleher}, R. and {Lelli}, F.},
        title = "{Galaxy clusters in Milgromian dynamics: Missing matter, hydrostatic bias, and the external field effect}",
      journal = {\aap},
         year = 2024,
        month = aug,
       volume = {688},
          eid = {A78},
        pages = {A78},
          doi = {10.1051/0004-6361/202449968},
archivePrefix = {arXiv},
       eprint = {2405.08557},
 primaryClass = {astro-ph.CO},
       adsurl = {https://ui.adsabs.harvard.edu/abs/2024A&A...688A..78K}
}

@ARTICLE{Cappellari2011,
       author = {{Cappellari}, Michele and {Emsellem}, Eric and {Krajnovi{\'c}}, Davor and {McDermid}, Richard M. and {Scott}, Nicholas and {Verdoes Kleijn}, G.~A. and {Young}, Lisa M. and {Alatalo}, Katherine and {Bacon}, R. and {Blitz}, Leo and {Bois}, Maxime and {Bournaud}, Fr{\'e}d{\'e}ric and {Bureau}, M. and {Davies}, Roger L. and {Davis}, Timothy A. and {de Zeeuw}, P.~T. and {Duc}, Pierre-Alain and {Khochfar}, Sadegh and {Kuntschner}, Harald and {Lablanche}, Pierre-Yves and {Morganti}, Raffaella and {Naab}, Thorsten and {Oosterloo}, Tom and {Sarzi}, Marc and {Serra}, Paolo and {Weijmans}, Anne-Marie},
        title = "{The ATLAS$^{3D}$ project - I. A volume-limited sample of 260 nearby early-type galaxies: science goals and selection criteria}",
      journal = {\mnras},
         year = 2011,
        month = may,
       volume = {413},
       number = {2},
        pages = {813-836},
          doi = {10.1111/j.1365-2966.2010.18174.x},
archivePrefix = {arXiv},
       eprint = {1012.1551},
 primaryClass = {astro-ph.CO},
       adsurl = {https://ui.adsabs.harvard.edu/abs/2011MNRAS.413..813C}
}

@ARTICLE{Cappellari2013,
       author = {{Cappellari}, Michele},
        title = "{Effect of Environment on Galaxies' Mass-Size Distribution: Unveiling the Transition from outside-in to inside-out Evolution}",
      journal = {\apjl},
         year = 2013,
        month = nov,
       volume = {778},
       number = {1},
          eid = {L2},
        pages = {L2},
          doi = {10.1088/2041-8205/778/1/L2},
archivePrefix = {arXiv},
       eprint = {1309.1136},
 primaryClass = {astro-ph.CO},
       adsurl = {https://ui.adsabs.harvard.edu/abs/2013ApJ...778L...2C}
}

@ARTICLE{Cappellari2013a,
       author = {{Cappellari}, Michele and {Scott}, Nicholas and {Alatalo}, Katherine and {Blitz}, Leo and {Bois}, Maxime and {Bournaud}, Fr{\'e}d{\'e}ric and {Bureau}, Martin and {Crocker}, Alison F. and {Davies}, Roger L. and {Davis}, Timothy A. and {de Zeeuw}, P.~T. and {Duc}, Pierre-Alain and {Emsellem}, Eric and {Khochfar}, Sadegh and {Krajnovi{\'c}}, Davor and {Kuntschner}, Harald and {McDermid}, Richard M. and {Morganti}, Raffaella and {Naab}, Thorsten and {Oosterloo}, Tom and {Sarzi}, Marc and {Serra}, Paolo and {Weijmans}, Anne-Marie and {Young}, Lisa M.},
        title = "{The ATLAS$^{3D}$ project - XV. A population of galaxies with rising mass-to-light ratios with radius: evidence for inside-out quenching?}",
      journal = {\mnras},
         year = 2013,
        month = may,
       volume = {432},
       number = {3},
        pages = {1709-1741},
          doi = {10.1093/mnras/stt562},
archivePrefix = {arXiv},
       eprint = {1208.3523},
 primaryClass = {astro-ph.CO},
       adsurl = {https://ui.adsabs.harvard.edu/abs/2013MNRAS.432.1709C}
}

@ARTICLE{Cappellari2013b,
       author = {{Cappellari}, Michele and {McDermid}, Richard M. and {Alatalo}, Katherine and {Blitz}, Leo and {Bois}, Maxime and {Bournaud}, Fr{\'e}d{\'e}ric and {Bureau}, M. and {Crocker}, Alison F. and {Davies}, Roger L. and {Davis}, Timothy A. and {de Zeeuw}, P.~T. and {Duc}, Pierre-Alain and {Emsellem}, Eric and {Khochfar}, Sadegh and {Krajnovi{\'c}}, Davor and {Kuntschner}, Harald and {Morganti}, Raffaella and {Naab}, Thorsten and {Oosterloo}, Tom and {Sarzi}, Marc and {Scott}, Nicholas and {Serra}, Paolo and {Weijmans}, Anne-Marie and {Young}, Lisa M.},
        title = "{The ATLAS$^{3D}$ project - XX. Mass-size and mass-{\ensuremath{\sigma}} distributions of early-type galaxies: bulge fraction drives kinematics, mass-to-light ratio, molecular gas fraction and stellar initial mass function}",
      journal = {\mnras},
         year = 2013,
        month = jul,
       volume = {432},
       number = {3},
        pages = {1862-1893},
          doi = {10.1093/mnras/stt644},
archivePrefix = {arXiv},
       eprint = {1208.3523},
 primaryClass = {astro-ph.CO},
       adsurl = {https://ui.adsabs.harvard.edu/abs/2013MNRAS.432.1862C}
}

@ARTICLE{Chae2021,
       author = {{Chae}, Kyu-Hyun and {Desmond}, Harry and {Lelli}, Federico and {McGaugh}, Stacy S. and {Schombert}, James M.},
        title = "{Testing the Strong Equivalence Principle. II. Relating the External Field Effect in Galaxy Rotation Curves to the Large-scale Structure of the Universe}",
      journal = {\apj},
         year = 2021,
        month = nov,
       volume = {921},
       number = {2},
          eid = {104},
        pages = {104},
          doi = {10.3847/1538-4357/ac1bba},
archivePrefix = {arXiv},
       eprint = {2109.04745},
 primaryClass = {astro-ph.GA},
       adsurl = {https://ui.adsabs.harvard.edu/abs/2021ApJ...921..104C}
}

@ARTICLE{Crain2015,
       author = {{Crain}, Robert A. and {Schaye}, Joop and {Bower}, Richard G. and {Furlong}, Michelle and {Schaller}, Matthieu and {Theuns}, Tom and {Dalla Vecchia}, Claudio and {Frenk}, Carlos S. and {McCarthy}, Ian G. and {Helly}, John C. and {Jenkins}, Adrian and {Rosas-Guevara}, Yetli M. and {White}, Simon D.~M. and {Trayford}, James W.},
        title = "{The EAGLE simulations of galaxy formation: calibration of subgrid physics and model variations}",
      journal = {\mnras},
         year = 2015,
        month = jun,
       volume = {450},
       number = {2},
        pages = {1937-1961},
          doi = {10.1093/mnras/stv725},
archivePrefix = {arXiv},
       eprint = {1501.01311},
 primaryClass = {astro-ph.GA},
       adsurl = {https://ui.adsabs.harvard.edu/abs/2015MNRAS.450.1937C}
}

@ARTICLE{Desmond2017,
       author = {{Desmond}, Harry and {Wechsler}, Risa H.},
        title = "{The Faber-Jackson relation and Fundamental Plane from halo abundance matching}",
      journal = {\mnras},
         year = 2017,
        month = feb,
       volume = {465},
       number = {1},
        pages = {820-833},
          doi = {10.1093/mnras/stw2804},
archivePrefix = {arXiv},
       eprint = {1604.04670},
 primaryClass = {astro-ph.GA},
       adsurl = {https://ui.adsabs.harvard.edu/abs/2017MNRAS.465..820D}
}

@ARTICLE{Dominguez2022,
       author = {{Dom{\'\i}nguez S{\'a}nchez}, H. and {Margalef}, B. and {Bernardi}, M. and {Huertas-Company}, M.},
        title = "{SDSS-IV DR17: final release of MaNGA PyMorph photometric and deep-learning morphological catalogues}",
      journal = {\mnras},
         year = 2022,
        month = jan,
       volume = {509},
       number = {3},
        pages = {4024-4036},
          doi = {10.1093/mnras/stab3089},
archivePrefix = {arXiv},
       eprint = {2110.10694},
 primaryClass = {astro-ph.GA},
       adsurl = {https://ui.adsabs.harvard.edu/abs/2022MNRAS.509.4024D}
}

@ARTICLE{Duann2023,
       author = {{Duann}, Yi and {Tian}, Yong and {Ko}, Chung-Ming},
        title = "{Classifying MaNGA velocity dispersion profiles by machine learning}",
      journal = {RAS Techniques and Instruments},
         year = 2023,
        month = jan,
       volume = {2},
       number = {1},
        pages = {649-656},
          doi = {10.1093/rasti/rzad044},
       adsurl = {https://ui.adsabs.harvard.edu/abs/2023RASTI...2..649D}
}

@ARTICLE{Djorgovski1987,
       author = {{Djorgovski}, S. and {Davis}, Marc},
        title = "{Fundamental Properties of Elliptical Galaxies}",
      journal = {\apj},
         year = 1987,
        month = feb,
       volume = {313},
        pages = {59},
          doi = {10.1086/164948},
       adsurl = {https://ui.adsabs.harvard.edu/abs/1987ApJ...313...59D}
}

@ARTICLE{Dressler1987,
       author = {{Dressler}, Alan and {Lynden-Bell}, Donald and {Burstein}, David and {Davies}, Roger L. and {Faber}, S.~M. and {Terlevich}, Roberto and {Wegner}, Gary},
        title = "{Spectroscopy and Photometry of Elliptical Galaxies. I. New Distance Estimator}",
      journal = {\apj},
         year = 1987,
        month = feb,
       volume = {313},
        pages = {42},
          doi = {10.1086/164947},
       adsurl = {https://ui.adsabs.harvard.edu/abs/1987ApJ...313...42D}
}

@ARTICLE{Ciotti1996,
       author = {{Ciotti}, L. and {Lanzoni}, B. and {Renzini}, A.},
        title = "{The tilt of the fundamental plane of elliptical galaxies - I. Exploring dynamical and structural effects}",
      journal = {\mnras},
         year = 1996,
        month = sep,
       volume = {282},
       number = {1},
        pages = {1-12},
          doi = {10.1093/mnras/282.1.1},
archivePrefix = {arXiv},
       eprint = {astro-ph/9601100},
 primaryClass = {astro-ph},
       adsurl = {https://ui.adsabs.harvard.edu/abs/1996MNRAS.282....1C}
}

@ARTICLE{Cappellari2006,
       author = {{Cappellari}, Michele and {Bacon}, R. and {Bureau}, M. and {Damen}, M.~C. and {Davies}, Roger L. and {de Zeeuw}, P.~T. and {Emsellem}, Eric and {Falc{\'o}n-Barroso}, Jes{\'u}s and {Krajnovi{\'c}}, Davor and {Kuntschner}, Harald and {McDermid}, Richard M. and {Peletier}, Reynier F. and {Sarzi}, Marc and {van den Bosch}, Remco C.~E. and {van de Ven}, Glenn},
        title = "{The SAURON project - IV. The mass-to-light ratio, the virial mass estimator and the Fundamental Plane of elliptical and lenticular galaxies}",
      journal = {\mnras},
         year = 2006,
        month = mar,
       volume = {366},
       number = {4},
        pages = {1126-1150},
          doi = {10.1111/j.1365-2966.2005.09981.x},
archivePrefix = {arXiv},
       eprint = {astro-ph/0505042},
 primaryClass = {astro-ph},
       adsurl = {https://ui.adsabs.harvard.edu/abs/2006MNRAS.366.1126C}
}

@ARTICLE{Bolton2007,
       author = {{Bolton}, Adam S. and {Burles}, Scott and {Treu}, Tommaso and {Koopmans}, L{\'e}on V.~E. and {Moustakas}, Leonidas A.},
        title = "{A More Fundamental Plane}",
      journal = {\apjl},
         year = 2007,
        month = aug,
       volume = {665},
       number = {2},
        pages = {L105-L108},
          doi = {10.1086/521357},
archivePrefix = {arXiv},
       eprint = {astro-ph/0701706},
 primaryClass = {astro-ph},
       adsurl = {https://ui.adsabs.harvard.edu/abs/2007ApJ...665L.105B}
}

@ARTICLE{Eftekhari2022,
       author = {{Eftekhari}, F. Sara and {Peletier}, Reynier F. and {Scott}, Nicholas and {Mieske}, Steffen and {Bland-Hawthorn}, Joss and {Bryant}, Julia J. and {Cantiello}, Michele and {Croom}, Scott M. and {Drinkwater}, Michael J. and {Falc{\'o}n-Barroso}, J{\'e}sus and {Hilker}, Michael and {Iodice}, Enrichetta and {Napolitano}, Nicola R. and {Spavone}, Marilena and {Valentijn}, Edwin A. and {van de Ven}, Glenn and {Venhola}, Aku},
        title = "{The SAMI-Fornax Dwarfs Survey - II. The Stellar Mass Fundamental Plane and the dark matter fraction of dwarf galaxies}",
      journal = {\mnras},
         year = 2022,
        month = dec,
       volume = {517},
       number = {4},
        pages = {4714-4735},
          doi = {10.1093/mnras/stac2606},
archivePrefix = {arXiv},
       eprint = {2209.05525},
 primaryClass = {astro-ph.GA},
       adsurl = {https://ui.adsabs.harvard.edu/abs/2022MNRAS.517.4714E}
}

@ARTICLE{Lelli2022,
       author = {{Lelli}, Federico},
        title = "{Gas dynamics in dwarf galaxies as testbeds for dark matter and galaxy evolution}",
      journal = {Nature Astronomy},
         year = 2022,
        month = jan,
       volume = {6},
        pages = {35-47},
          doi = {10.1038/s41550-021-01562-2},
archivePrefix = {arXiv},
       eprint = {2201.11752},
 primaryClass = {astro-ph.GA},
       adsurl = {https://ui.adsabs.harvard.edu/abs/2022NatAs...6...35L}
}

@ARTICLE{FJ1976,
       author = {{Faber}, S.~M. and {Jackson}, R.~E.},
        title = "{Velocity dispersions and mass-to-light ratios for elliptical galaxies.}",
      journal = {\apj},
         year = 1976,
        month = mar,
       volume = {204},
        pages = {668-683},
          doi = {10.1086/154215},
       adsurl = {https://ui.adsabs.harvard.edu/abs/1976ApJ...204..668F}
}

@ARTICLE{FM2012,
       author = {{Famaey}, Beno{\^\i}t and {McGaugh}, Stacy S.},
        title = "{Modified Newtonian Dynamics (MOND): Observational Phenomenology and Relativistic Extensions}",
      journal = {Living Reviews in Relativity},
         year = 2012,
        month = dec,
       volume = {15},
       number = {1},
          eid = {10},
        pages = {10},
          doi = {10.12942/lrr-2012-10},
archivePrefix = {arXiv},
       eprint = {1112.3960},
 primaryClass = {astro-ph.CO},
       adsurl = {https://ui.adsabs.harvard.edu/abs/2012LRR....15...10F}
}

@ARTICLE{Krajnovic2011,
       author = {{Krajnovi{\'c}}, Davor and {Emsellem}, Eric and {Cappellari}, Michele and {Alatalo}, Katherine and {Blitz}, Leo and {Bois}, Maxime and {Bournaud}, Fr{\'e}d{\'e}ric and {Bureau}, Martin and {Davies}, Roger L. and {Davis}, Timothy A. and {de Zeeuw}, P.~T. and {Khochfar}, Sadegh and {Kuntschner}, Harald and {Lablanche}, Pierre-Yves and {McDermid}, Richard M. and {Morganti}, Raffaella and {Naab}, Thorsten and {Oosterloo}, Tom and {Sarzi}, Marc and {Scott}, Nicholas and {Serra}, Paolo and {Weijmans}, Anne-Marie and {Young}, Lisa M.},
        title = "{The ATLAS$^{3D}$ project - II. Morphologies, kinemetric features and alignment between photometric and kinematic axes of early-type galaxies}",
      journal = {\mnras},
         year = 2011,
        month = jul,
       volume = {414},
       number = {4},
        pages = {2923-2949},
          doi = {10.1111/j.1365-2966.2011.18560.x},
archivePrefix = {arXiv},
       eprint = {1102.3801},
 primaryClass = {astro-ph.CO},
       adsurl = {https://ui.adsabs.harvard.edu/abs/2011MNRAS.414.2923K}
}

@ARTICLE{Krajnovic2013,
       author = {{Krajnovi{\'c}}, Davor and {Alatalo}, Katherine and {Blitz}, Leo and {Bois}, Maxime and {Bournaud}, Fr{\'e}d{\'e}ric and {Bureau}, Martin and {Cappellari}, Michele and {Davies}, Roger L. and {Davis}, Timothy A. and {de Zeeuw}, P.~T. and {Duc}, Pierre-Alain and {Emsellem}, Eric and {Khochfar}, Sadegh and {Kuntschner}, Harald and {McDermid}, Richard M. and {Morganti}, Raffaella and {Naab}, Thorsten and {Oosterloo}, Tom and {Sarzi}, Marc and {Scott}, Nicholas and {Serra}, Paolo and {Weijmans}, Anne-Marie and {Young}, Lisa M.},
        title = "{The ATLAS$^{3D}$ project - XVII. Linking photometric and kinematic signatures of stellar discs in early-type galaxies}",
      journal = {\mnras},
         year = 2013,
        month = jul,
       volume = {432},
       number = {3},
        pages = {1768-1795},
          doi = {10.1093/mnras/sts315},
archivePrefix = {arXiv},
       eprint = {1210.8167},
 primaryClass = {astro-ph.CO},
       adsurl = {https://ui.adsabs.harvard.edu/abs/2013MNRAS.432.1768K}
}

@ARTICLE{Kauffmann2003,
       author = {{Kauffmann}, Guinevere and {Heckman}, Timothy M. and {White}, Simon D.~M. and {Charlot}, St{\'e}phane and {Tremonti}, Christy and {Brinchmann}, Jarle and {Bruzual}, Gustavo and {Peng}, Eric W. and {Seibert}, Mark and {Bernardi}, Mariangela and {Blanton}, Michael and {Brinkmann}, Jon and {Castander}, Francisco and {Cs{\'a}bai}, Istvan and {Fukugita}, Masataka and {Ivezic}, Zeljko and {Munn}, Jeffrey A. and {Nichol}, Robert C. and {Padmanabhan}, Nikhil and {Thakar}, Aniruddha R. and {Weinberg}, David H. and {York}, Donald},
        title = "{Stellar masses and star formation histories for {}10$^{5}$ galaxies from the Sloan Digital Sky Survey}",
      journal = {\mnras},
         year = 2003,
        month = may,
       volume = {341},
       number = {1},
        pages = {33-53},
          doi = {10.1046/j.1365-8711.2003.06291.x},
archivePrefix = {arXiv},
       eprint = {astro-ph/0204055},
 primaryClass = {astro-ph},
       adsurl = {https://ui.adsabs.harvard.edu/abs/2003MNRAS.341...33K}
}

@ARTICLE{Kroupa2001,
       author = {{Kroupa}, Pavel},
        title = "{On the variation of the initial mass function}",
      journal = {\mnras},
         year = 2001,
        month = apr,
       volume = {322},
       number = {2},
        pages = {231-246},
          doi = {10.1046/j.1365-8711.2001.04022.x},
archivePrefix = {arXiv},
       eprint = {astro-ph/0009005},
 primaryClass = {astro-ph},
       adsurl = {https://ui.adsabs.harvard.edu/abs/2001MNRAS.322..231K}
}

@ARTICLE{Lelli2016,
       author = {{Lelli}, Federico and {McGaugh}, Stacy S. and {Schombert}, James M.},
        title = "{The Small Scatter of the Baryonic Tully-Fisher Relation}",
      journal = {\apjl},
         year = 2016,
        month = jan,
       volume = {816},
       number = {1},
          eid = {L14},
        pages = {L14},
          doi = {10.3847/2041-8205/816/1/L14},
archivePrefix = {arXiv},
       eprint = {1512.04543},
 primaryClass = {astro-ph.GA},
       adsurl = {https://ui.adsabs.harvard.edu/abs/2016ApJ...816L..14L}
}

@ARTICLE{Lelli2017,
       author = {{Lelli}, Federico and {McGaugh}, Stacy S. and {Schombert}, James M. and {Pawlowski}, Marcel S.},
        title = "{One Law to Rule Them All: The Radial Acceleration Relation of Galaxies}",
      journal = {\apj},
         year = 2017,
        month = feb,
       volume = {836},
       number = {2},
          eid = {152},
        pages = {152},
          doi = {10.3847/1538-4357/836/2/152},
archivePrefix = {arXiv},
       eprint = {1610.08981},
 primaryClass = {astro-ph.GA},
       adsurl = {https://ui.adsabs.harvard.edu/abs/2017ApJ...836..152L}
}

@ARTICLE{Lelli2019,
       author = {{Lelli}, Federico and {McGaugh}, Stacy S. and {Schombert}, James M. and {Desmond}, Harry and {Katz}, Harley},
        title = "{The baryonic Tully-Fisher relation for different velocity definitions and implications for galaxy angular momentum}",
      journal = {\mnras},
         year = 2019,
        month = apr,
       volume = {484},
       number = {3},
        pages = {3267-3278},
          doi = {10.1093/mnras/stz205},
archivePrefix = {arXiv},
       eprint = {1901.05966},
 primaryClass = {astro-ph.GA},
       adsurl = {https://ui.adsabs.harvard.edu/abs/2019MNRAS.484.3267L}
}

@ARTICLE{McCarthy2017,
       author = {{McCarthy}, Ian G. and {Schaye}, Joop and {Bird}, Simeon and {Le Brun}, Amandine M.~C.},
        title = "{The BAHAMAS project: calibrated hydrodynamical simulations for large-scale structure cosmology}",
      journal = {\mnras},
         year = 2017,
        month = mar,
       volume = {465},
       number = {3},
        pages = {2936-2965},
          doi = {10.1093/mnras/stw2792},
archivePrefix = {arXiv},
       eprint = {1603.02702},
 primaryClass = {astro-ph.CO},
       adsurl = {https://ui.adsabs.harvard.edu/abs/2017MNRAS.465.2936M}
}

@ARTICLE{McGaugh2000,
       author = {{McGaugh}, S.~S. and {Schombert}, J.~M. and {Bothun}, G.~D. and {de Blok}, W.~J.~G.},
        title = "{The Baryonic Tully-Fisher Relation}",
      journal = {\apjl},
         year = 2000,
        month = apr,
       volume = {533},
       number = {2},
        pages = {L99-L102},
          doi = {10.1086/312628},
archivePrefix = {arXiv},
       eprint = {astro-ph/0003001},
 primaryClass = {astro-ph},
       adsurl = {https://ui.adsabs.harvard.edu/abs/2000ApJ...533L..99M}
}

@ARTICLE{Milgrom1983,
       author = {{Milgrom}, M.},
        title = "{A modification of the Newtonian dynamics as a possible alternative to the hidden mass hypothesis.}",
      journal = {\apj},
         year = 1983,
        month = jul,
       volume = {270},
        pages = {365-370},
          doi = {10.1086/161130},
       adsurl = {https://ui.adsabs.harvard.edu/abs/1983ApJ...270..365M}
}

@ARTICLE{Milgrom1984,
       author = {{Milgrom}, M.},
        title = "{Isothermal spheres in the modified dynamics}",
      journal = {\apj},
         year = 1984,
        month = dec,
       volume = {287},
        pages = {571-576},
          doi = {10.1086/162716},
       adsurl = {https://ui.adsabs.harvard.edu/abs/1984ApJ...287..571M}
}

@ARTICLE{Milgrom2010,
       author = {{Milgrom}, Mordehai},
        title = "{Quasi-linear formulation of MOND}",
      journal = {\mnras},
         year = 2010,
        month = apr,
       volume = {403},
       number = {2},
        pages = {886-895},
          doi = {10.1111/j.1365-2966.2009.16184.x},
archivePrefix = {arXiv},
       eprint = {0911.5464},
 primaryClass = {astro-ph.CO},
       adsurl = {https://ui.adsabs.harvard.edu/abs/2010MNRAS.403..886M}
}

@ARTICLE{Milgrom2014a,
       author = {{Milgrom}, Mordehai},
        title = "{General virial theorem for modified-gravity MOND}",
      journal = {\prd},
         year = 2014,
        month = jan,
       volume = {89},
       number = {2},
          eid = {024016},
        pages = {024016},
          doi = {10.1103/PhysRevD.89.024016},
archivePrefix = {arXiv},
       eprint = {1311.2579},
 primaryClass = {gr-qc},
       adsurl = {https://ui.adsabs.harvard.edu/abs/2014PhRvD..89b4016M}
}

@ARTICLE{Milgrom2014b,
       author = {{Milgrom}, Mordehai},
        title = "{MOND laws of galactic dynamics}",
      journal = {\mnras},
         year = 2014,
        month = jan,
       volume = {437},
       number = {3},
        pages = {2531-2541},
          doi = {10.1093/mnras/stt2066},
archivePrefix = {arXiv},
       eprint = {1212.2568},
 primaryClass = {astro-ph.CO},
       adsurl = {https://ui.adsabs.harvard.edu/abs/2014MNRAS.437.2531M}
}

@ARTICLE{Milgrom2019,
       author = {{Milgrom}, Mordehai},
        title = "{MOND in galaxy groups: A superior sample}",
      journal = {\prd},
         year = 2019,
        month = feb,
       volume = {99},
       number = {4},
          eid = {044041},
        pages = {044041},
          doi = {10.1103/PhysRevD.99.044041},
archivePrefix = {arXiv},
       eprint = {1811.12233},
 primaryClass = {astro-ph.GA},
       adsurl = {https://ui.adsabs.harvard.edu/abs/2019PhRvD..99d4041M}
}

@ARTICLE{Milgrom2025,
       author = {{Milgrom}, Mordehai},
        title = "{The deep-MOND limit -- a study in Primary vs secondary predictions}",
      journal = {arXiv e-prints},
         year = 2025,
        month = oct,
          eid = {arXiv:2510.16520},
        pages = {arXiv:2510.16520},
          doi = {10.48550/arXiv.2510.16520},
archivePrefix = {arXiv},
       eprint = {2510.16520},
 primaryClass = {gr-qc},
       adsurl = {https://ui.adsabs.harvard.edu/abs/2025arXiv251016520M}
}

@ARTICLE{MK2011,
       author = {{Makarov}, Dmitry and {Karachentsev}, Igor},
        title = "{Galaxy groups and clouds in the local (z{\ensuremath{\sim}} 0.01) Universe}",
      journal = {\mnras},
         year = 2011,
        month = apr,
       volume = {412},
       number = {4},
        pages = {2498-2520},
          doi = {10.1111/j.1365-2966.2010.18071.x},
archivePrefix = {arXiv},
       eprint = {1011.6277},
 primaryClass = {astro-ph.CO},
       adsurl = {https://ui.adsabs.harvard.edu/abs/2011MNRAS.412.2498M}
}

@ARTICLE{Nelson2019,
       author = {{Nelson}, Dylan and {Springel}, Volker and {Pillepich}, Annalisa and {Rodriguez-Gomez}, Vicente and {Torrey}, Paul and {Genel}, Shy and {Vogelsberger}, Mark and {Pakmor}, Ruediger and {Marinacci}, Federico and {Weinberger}, Rainer and {Kelley}, Luke and {Lovell}, Mark and {Diemer}, Benedikt and {Hernquist}, Lars},
        title = "{The IllustrisTNG simulations: public data release}",
      journal = {Computational Astrophysics and Cosmology},
         year = 2019,
        month = may,
       volume = {6},
       number = {1},
          eid = {2},
        pages = {2},
          doi = {10.1186/s40668-019-0028-x},
archivePrefix = {arXiv},
       eprint = {1812.05609},
 primaryClass = {astro-ph.GA},
       adsurl = {https://ui.adsabs.harvard.edu/abs/2019ComAC...6....2N}
}

@ARTICLE{Pillepich2018,
       author = {{Pillepich}, Annalisa and {Nelson}, Dylan and {Hernquist}, Lars and {Springel}, Volker and {Pakmor}, R{\"u}diger and {Torrey}, Paul and {Weinberger}, Rainer and {Genel}, Shy and {Naiman}, Jill P. and {Marinacci}, Federico and {Vogelsberger}, Mark},
        title = "{First results from the IllustrisTNG simulations: the stellar mass content of groups and clusters of galaxies}",
      journal = {\mnras},
         year = 2018,
        month = mar,
       volume = {475},
       number = {1},
        pages = {648-675},
          doi = {10.1093/mnras/stx3112},
archivePrefix = {arXiv},
       eprint = {1707.03406},
 primaryClass = {astro-ph.GA},
       adsurl = {https://ui.adsabs.harvard.edu/abs/2018MNRAS.475..648P}
}

@ARTICLE{ST2024,
       author = {{Sadhu}, Pradyumna and {Tian}, Yong},
        title = "{Examining baryonic Faber-Jackson relation in galaxy groups}",
      journal = {\mnras},
         year = 2024,
        month = mar,
       volume = {528},
       number = {4},
        pages = {5612-5623},
          doi = {10.1093/mnras/stae343},
archivePrefix = {arXiv},
       eprint = {2402.13320},
 primaryClass = {astro-ph.CO},
       adsurl = {https://ui.adsabs.harvard.edu/abs/2024MNRAS.528.5612S}
}

@ARTICLE{Sanders2010,
       author = {{Sanders}, R.~H.},
        title = "{The universal Faber-Jackson relation}",
      journal = {\mnras},
         year = 2010,
        month = sep,
       volume = {407},
       number = {2},
        pages = {1128-1134},
          doi = {10.1111/j.1365-2966.2010.16957.x},
archivePrefix = {arXiv},
       eprint = {1002.2765},
 primaryClass = {astro-ph.CO},
       adsurl = {https://ui.adsabs.harvard.edu/abs/2010MNRAS.407.1128S}
}

@ARTICLE{Serra2012,
       author = {{Serra}, Paolo and {Oosterloo}, Tom and {Morganti}, Raffaella and {Alatalo}, Katherine and {Blitz}, Leo and {Bois}, Maxime and {Bournaud}, Fr{\'e}d{\'e}ric and {Bureau}, Martin and {Cappellari}, Michele and {Crocker}, Alison F. and {Davies}, Roger L. and {Davis}, Timothy A. and {de Zeeuw}, P.~T. and {Duc}, Pierre-Alain and {Emsellem}, Eric and {Khochfar}, Sadegh and {Krajnovi{\'c}}, Davor and {Kuntschner}, Harald and {Lablanche}, Pierre-Yves and {McDermid}, Richard M. and {Naab}, Thorsten and {Sarzi}, Marc and {Scott}, Nicholas and {Trager}, Scott C. and {Weijmans}, Anne-Marie and {Young}, Lisa M.},
        title = "{The ATLAS$^{3D}$ project - XIII. Mass and morphology of H I in early-type galaxies as a function of environment}",
      journal = {\mnras},
         year = 2012,
        month = may,
       volume = {422},
       number = {3},
        pages = {1835-1862},
          doi = {10.1111/j.1365-2966.2012.20219.x},
archivePrefix = {arXiv},
       eprint = {1111.4241},
 primaryClass = {astro-ph.CO},
       adsurl = {https://ui.adsabs.harvard.edu/abs/2012MNRAS.422.1835S}
}

@ARTICLE{Scott2020,
       author = {{Scott}, Nicholas and {Eftekhari}, F. Sara and {Peletier}, Reynier F. and {Bryant}, Julia J. and {Bland-Hawthorn}, Joss and {Capaccioli}, Massimo and {Croom}, Scott M. and {Drinkwater}, Michael and {Falc{\'o}n-Barroso}, J{\'e}sus and {Hilker}, Michael and {Iodice}, Enrichetta and {Lorente}, Nuria F.~P. and {Mieske}, Steffen and {Spavone}, Marilena and {van de Ven}, Glenn and {Venhola}, Aku},
        title = "{The SAMI-Fornax Dwarfs Survey I: sample, observations, and the specific stellar angular momentum of dwarf elliptical galaxies}",
      journal = {\mnras},
         year = 2020,
        month = sep,
       volume = {497},
       number = {2},
        pages = {1571-1582},
          doi = {10.1093/mnras/staa2042},
archivePrefix = {arXiv},
       eprint = {2007.04492},
 primaryClass = {astro-ph.GA},
       adsurl = {https://ui.adsabs.harvard.edu/abs/2020MNRAS.497.1571S}
}

@ARTICLE{Taylor2011,
       author = {{Taylor}, Edward N. and {Hopkins}, Andrew M. and {Baldry}, Ivan K. and {Brown}, Michael J.~I. and {Driver}, Simon P. and {Kelvin}, Lee S. and {Hill}, David T. and {Robotham}, Aaron S.~G. and {Bland-Hawthorn}, Joss and {Jones}, D.~H. and {Sharp}, R.~G. and {Thomas}, Daniel and {Liske}, Jochen and {Loveday}, Jon and {Norberg}, Peder and {Peacock}, J.~A. and {Bamford}, Steven P. and {Brough}, Sarah and {Colless}, Matthew and {Cameron}, Ewan and {Conselice}, Christopher J. and {Croom}, Scott M. and {Frenk}, C.~S. and {Gunawardhana}, Madusha and {Kuijken}, Konrad and {Nichol}, R.~C. and {Parkinson}, H.~R. and {Phillipps}, S. and {Pimbblet}, K.~A. and {Popescu}, C.~C. and {Prescott}, Matthew and {Sutherland}, W.~J. and {Tuffs}, R.~J. and {van Kampen}, Eelco and {Wijesinghe}, D.},
        title = "{Galaxy And Mass Assembly (GAMA): stellar mass estimates}",
      journal = {\mnras},
         year = 2011,
        month = dec,
       volume = {418},
       number = {3},
        pages = {1587-1620},
          doi = {10.1111/j.1365-2966.2011.19536.x},
archivePrefix = {arXiv},
       eprint = {1108.0635},
 primaryClass = {astro-ph.CO},
       adsurl = {https://ui.adsabs.harvard.edu/abs/2011MNRAS.418.1587T}
}

@ARTICLE{Tian2021a,
       author = {{Tian}, Yong and {Cheng}, Han and {McGaugh}, Stacy S. and {Ko}, Chung-Ming and {Hsu}, Yun-Hsin},
        title = "{Mass-Velocity Dispersion Relation in MaNGA Brightest Cluster Galaxies}",
      journal = {\apjl},
         year = 2021,
        month = aug,
       volume = {917},
       number = {2},
          eid = {L24},
        pages = {L24},
          doi = {10.3847/2041-8213/ac1a18},
archivePrefix = {arXiv},
       eprint = {2108.08980},
 primaryClass = {astro-ph.GA},
       adsurl = {https://ui.adsabs.harvard.edu/abs/2021ApJ...917L..24T}
}

@ARTICLE{Toloba2014,
       author = {{Toloba}, E. and {Guhathakurta}, P. and {Peletier}, R.~F. and {Boselli}, A. and {Lisker}, T. and {Falc{\'o}n-Barroso}, J. and {Simon}, J.~D. and {van de Ven}, G. and {Paudel}, S. and {Emsellem}, E. and {Janz}, J. and {den Brok}, M. and {Gorgas}, J. and {Hensler}, G. and {Laurikainen}, E. and {Niemi}, S. -M. and {Ry{\'s}}, A. and {Salo}, H.},
        title = "{Stellar Kinematics and Structural Properties of Virgo Cluster Dwarf Early-type Galaxies from the SMAKCED Project. II. The Survey and a Systematic Analysis of Kinematic Anomalies and Asymmetries}",
      journal = {\apjs},
         year = 2014,
        month = dec,
       volume = {215},
       number = {2},
          eid = {17},
        pages = {17},
          doi = {10.1088/0067-0049/215/2/17},
archivePrefix = {arXiv},
       eprint = {1410.1550},
 primaryClass = {astro-ph.GA},
       adsurl = {https://ui.adsabs.harvard.edu/abs/2014ApJS..215...17T}
}

@ARTICLE{Vazdekis2010,
       author = {{Vazdekis}, A. and {S{\'a}nchez-Bl{\'a}zquez}, P. and {Falc{\'o}n-Barroso}, J. and {Cenarro}, A.~J. and {Beasley}, M.~A. and {Cardiel}, N. and {Gorgas}, J. and {Peletier}, R.~F.},
        title = "{Evolutionary stellar population synthesis with MILES - I. The base models and a new line index system}",
      journal = {\mnras},
         year = 2010,
        month = jun,
       volume = {404},
       number = {4},
        pages = {1639-1671},
          doi = {10.1111/j.1365-2966.2010.16407.x},
archivePrefix = {arXiv},
       eprint = {1004.4439},
 primaryClass = {astro-ph.CO},
       adsurl = {https://ui.adsabs.harvard.edu/abs/2010MNRAS.404.1639V}
}

%%%%%%%%%%%%%%%%%%%%%%%%%%%%%%%%%%%%%%%%%%%%%%%%%%%%%%%%%
%
%                Appendices
%
%%%%%%%%%%%%%%%%%%%%%%%%%%%%%%%%%%%%%%%%%%%%%%%%%%%%%%%%%
\clearpage

\appendix
\section{Samples}\label{app:samples}

\begin{figure}[!htb]
  \centering
  \includegraphics[width=0.95\columnwidth]{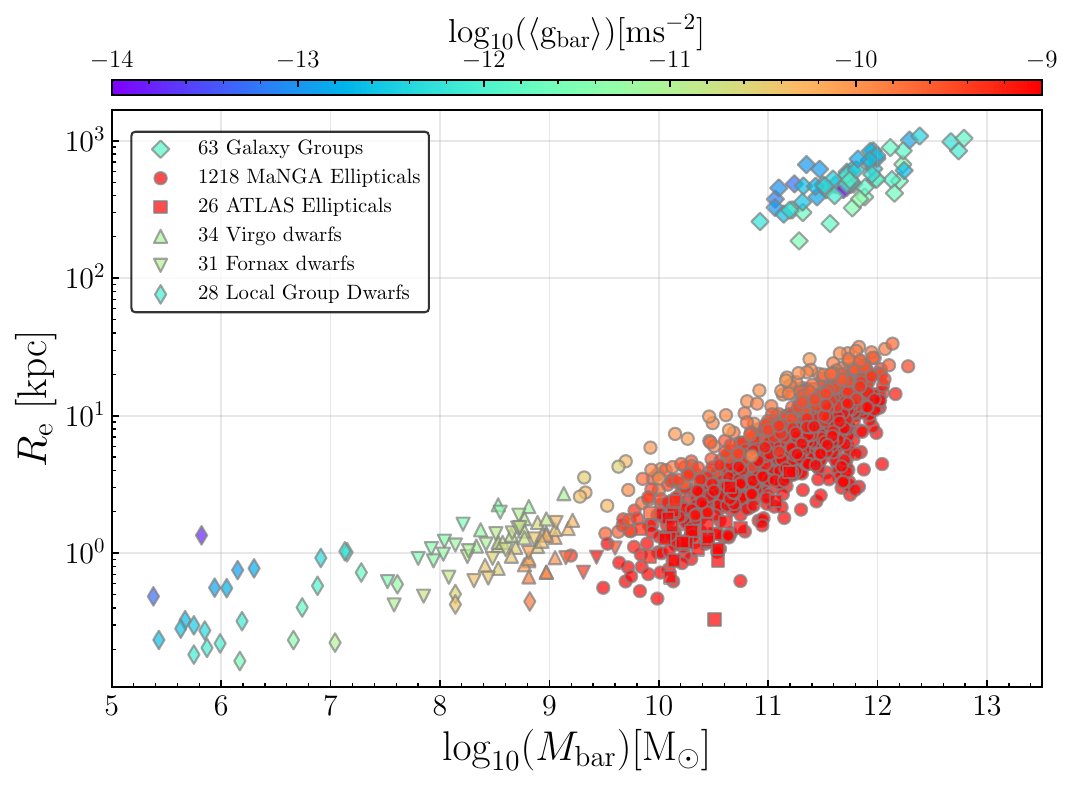}
  \caption{The relation between baryonic mass and characteristic size (effective radius for galaxies, mean radius for galaxy groups) of our sample. Different symbols represent different datasets: galaxy groups (large diamonds), MaNGA ellipticals (circles), ATLAS$^{3\rm D}$ ellipticals (squares), Virgo dwarfs (upward triangles), Fornax dwarfs (downward triangles), and Local Group dwarfs (small diamonds). The color coding represents the median baryonic gravitational acceleration, $\log_{10}(\gbarmed)$, within the effective radius (or the mean radius for galaxy groups)}.
  \label{fig:Data}
\end{figure}

The data analyzed in this paper are drawn from six public catalogs. A description of each catalog is provided below. Fig.~\ref{fig:Data} shows the baryonic mass against the effective radius (or the equivalent mean radius for galaxy groups). Our sample spans $\sim$8 dex in baryonic mass ($M_{\rm bar}\simeq10^5-10^{13}$ M$_\odot$) and $\sim$4 dex in characteristic size (from $\sim$100 pc to $\sim$1 Mpc). The complete data table is available at the CDS.

\paragraph{Galaxy Groups:}
We use data for 13 galaxy groups from~\citet{ST2024} and 50 groups from~\citet{Milgrom2019}, with original measurements primarily based on
\citet{MK2011}. The baryonic mass is given by the stellar mass of the member galaxies plus the hot gas mass from X-ray observations. Stellar masses are estimated from the $K$-band absolute magnitude ($M_{K}$) of each member galaxy using the relation from~\citet{Cappellari2013}, as adopted in~\citet{ST2024}:
\begin{equation}
\log_{10} \Mstar = 10.58 - 0.44(M_{K} + 23)\,.
\end{equation}
This relation is consistent with a Kroupa IMF. According to these measurements, the stellar mass dominates over the hot gas mass. In principle, there may be a warm-hot intergalactic medium (with temperatures of $10^5-10^6$ K) that is not observed in the X-rays. Given the considerable uncertainties on the amount of such warm-hot gas component, we do not include it in our baryonic mass estimate. Velocity dispersions are calculated using a robust biweight estimator applied to the line-of-sight velocities of group members within the mean radius, defined as the average projected radius of all member galaxies.

In the $\Lambda$CDM context, there should be missing baryons in galaxy groups, possibly in the form of warm-hot gas \citep[e.g.,][]{McGaugh26}. This does not need to hold in other paradigms, such as MOND. In this work, we consider only the baryonic mass that is directly observed: stars and X-ray gas. A substantial amount of missing baryons in galaxy groups will lead to a steeper slope and smaller intercept of the low-acceleration BFJR. It will also move galaxy groups close to the Newtonian expectation in the baryonic FP.

\paragraph{MaNGA Elliptical Galaxies:}
We selected galaxies from the Sloan Digital Sky Survey IV (SDSS-IV) MaNGA survey~\citep{Bundy2015}.
MaNGA provides two-dimensional spectroscopic maps for nearly 10,000 galaxies, enabling precise measurements of stellar kinematics. From this dataset, 2,632 elliptical galaxies were identified based on morphological classifications using a convolutional neural network, as presented in~\citet{Dominguez2022}.  To ensure that the velocity dispersion is a faithful tracer of the equilibrium gravitational potential, we excluded galaxies exhibiting ascending (5.6\percent) or irregular (2.1\percent) velocity dispersion profiles, as classified by~\citet{Duann2023}. Only galaxies with declining or flat velocity dispersion profiles were included in this study. Finally, we restrict to galaxies with velocity dispersion measurements out to at least one effective radius ($R_{\rm e}$), so we can calculate $\sigma_{\rm e}$ (the average line-of-sight velocity dispersion within $R_{\rm e}$). Our final sample comprises 1218 elliptical galaxies. The effective radius and \Sersic\ index are provided by the MaNGA PyMorph photometric Value Added Catalogue (MPP-VAC-DR17) in~\citet{Dominguez2022}.

Stellar masses are derived from fitting the spectral energy distribution (SED) assuming a Kroupa initial mass function~\citep[IMF,][]{Kauffmann2003, Kroupa2001}. The hot gas mass in elliptical galaxies was estimated using the scaling relation from~\citet{Chae2021}:
\begin{equation}
    \log(M_{\rm g,hot}/\Msun)=1.47\log(\Mstar/\Msun)-5.414\,.
    \label{eq:hotgas}
\end{equation}
Given that the cold gas mass of ellipticals may be (at most) a few percent of the stellar mass~\citep{Serra2012}, we assume $\Mbar \approx \Mstar + M_{\rm g,hot}.$

\paragraph{ATLAS$^{3\rm D}$ Elliptical Galaxies:}
This sample comprises 260 local early-type galaxies from the ATLAS$^{3\rm D}$ survey~\citep{Cappellari2011}. We restrict the sample to 68 elliptical galaxies using the morphological classifications from~\citet{Krajnovic2011}. For consistency with other samples, we limit our analysis to 26 galaxies with data extending out to at least one effective radius, enabling measurements of $\sigma_{\rm e}$.
Stellar masses and kinematic data are taken from~\citet{Cappellari2013a}, while the \Sersic\ indices and effective radii are adopted from~\citet{Krajnovic2013}.

Stellar masses are estimated using the mass-to-light ratio ($M/L$) at the effective radius, as determined with the Salpeter IMF in~\citet{Cappellari2013b}. To ensure consistency with other samples, we convert these values to a Kroupa IMF using $\Mstar = M_{\rm Salp}/1.6$. The hot gas mass is estimated using Eq.~(\ref{eq:hotgas}), yielding a median value of 25\percent\ of $\Mstar$ in our sample. Cold gas masses, as measured by~\citet{Serra2012}, contribute only a few percent relative to $\Mstar$. The total baryonic mass is thus defined as $\Mbar = \Mstar + \Mgas + M_{\rm g,hot}$.

\paragraph{Virgo Dwarfs:}
We selected 34 out of 39 Virgo dwarf galaxies from~\citet{Toloba2014}, requiring that the velocity dispersion within the effective radius exceed the rotational velocity. Stellar masses are estimated assuming a constant mass-to-light ratio of $\mathrm{M}/\mathrm{L}=0.73$ in the $H$-band at one effective radius by adopting the Kroupa IMF~\citep{Vazdekis2010}. The \Sersic\ index ranges from 1.0 to 2.2, as reported in~\citet{Toloba2014}; for a few galaxies without measured indices, we adopt $n = 1$ for the calculation of baryonic acceleration. Given their cluster environment, these dwarfs contain minimal gas, so we assume $\Mbar \simeq \Mstar$.

\paragraph{Fornax Dwarfs:}
Data for dwarf galaxies in Fornax are drawn from the SAMI-Fornax Dwarf Galaxy Survey~\citep{Scott2020, Eftekhari2022}. We selected 31 galaxies morphologically classified as dwarf ellipticals (dE). The SAMI integral-field spectrograph provides kinematic data extending beyond the half-light radius. The velocity dispersion is consistently measured within one effective radius. The \Sersic\ parameters used for the baryonic acceleration calculations are reported in Table 1 of~\citet{Eftekhari2022}. Stellar masses are estimated by \citet{Taylor2011, Eftekhari2022} using the Chabrier IMF, based on $r$-band measurements and $g-i$ and $r-i$ colors:
\begin{equation}
\log(\Mstar/\Msun)_{e} = 1.15 + 0.75(g-i) - 0.4M_{r,e} + 0.4(r-i)\,.
\end{equation}
The Chabrier IMF is virtually equivalent to the Kroupa IMF assumed for the other samples. Given their cluster environment, these dwarfs contain very little gas, so we assume $\Mbar \approx \Mstar$. 

\paragraph{Local Group Dwarf Spheroidals:}
To extend the BFJR to the lowest mass regime, we include dwarf spheroidal (dSph) satellite galaxies of the Milky Way and Andromeda. Data are taken from the compilation by~\citet{Lelli2017}. 
We select 28 dSph galaxies with luminosities greater than $10^5\,\Msun$ (so we do not consider the so-called "ultra-faint dwarfs" that have substantially more uncertain data) and with velocity dispersions measured from more than 20 member stars. Stellar masses are estimated assuming a mass-to-light ratio of $M_\star/L = 2$ in the $V$-band (for an Kroupa IMF), and a \Sersic\ index of $n = 1$ is adopted for all systems. As these satellites are almost entirely devoid of gas, their baryonic mass is effectively equal to their stellar mass.

Finally, we note that compiling data across eight orders of magnitude in mass inherently requires combining systems with different observational constraints. Consequently, the adopted velocity dispersions are not strictly identical observables across all subsamples. For individual elliptical and dwarf galaxies, we use the stellar line-of-sight velocity dispersion integrated within one effective radius ($\sigma_e$). In contrast, for galaxy groups, we use the velocity dispersion derived from the discrete line-of-sight velocities of member galaxies. 
While our homogenization is intended to provide a characteristic measure of pressure support across vastly different classes of systems, these quantities represent physically distinct tracers. These systematic differences between subsamples should be kept in mind when interpreting the absolute parameters of the scaling relations, as they naturally contribute additional intrinsic scatter to the overarching empirical trend.

\section{The baryonic Faber-Jackson relation for high-acceleration subsamples}\label{app:High_acc}

To characterize the BFJR in the high-acceleration regime, we fit a linear relation to sub-samples with $\gbarmed> X \aZero$, where $X$ is a number from 1 to 50. Similarly to Sect.\,\ref{sec:results}, we perform both orthogonal and vertical MCMC fits, explicitly accounting for measurement uncertainties in $\log_{10}(\Mbar)$ and $\log_{10}(\sigma_{\rm e})$, as well as intrinsic scatter. This methodology provides robust estimates of the slope and intercept even when measurement errors affect both axes. 

\begin{figure*}[t]
\centering
\includegraphics[width=0.6\columnwidth]{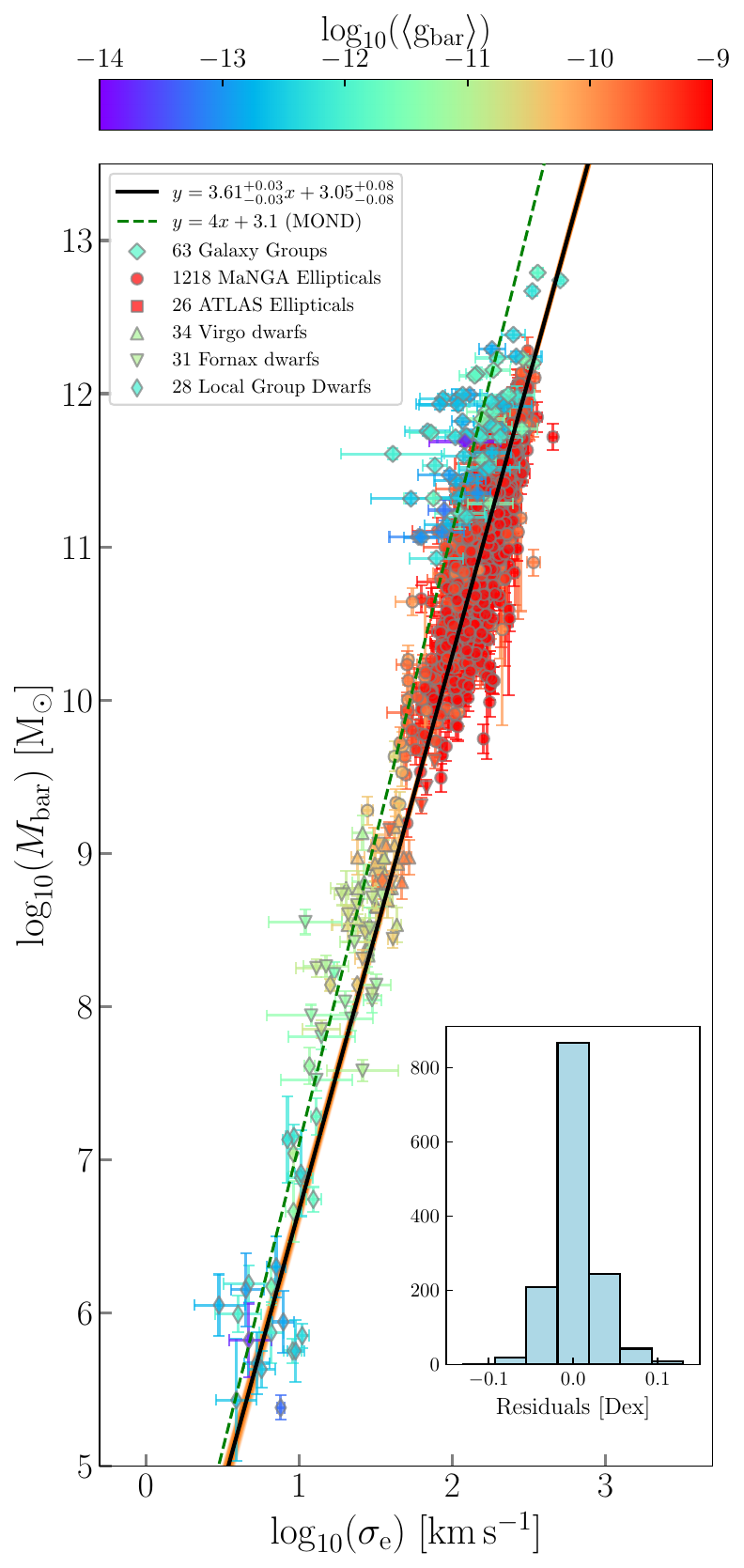}
\includegraphics[width=0.6\columnwidth]{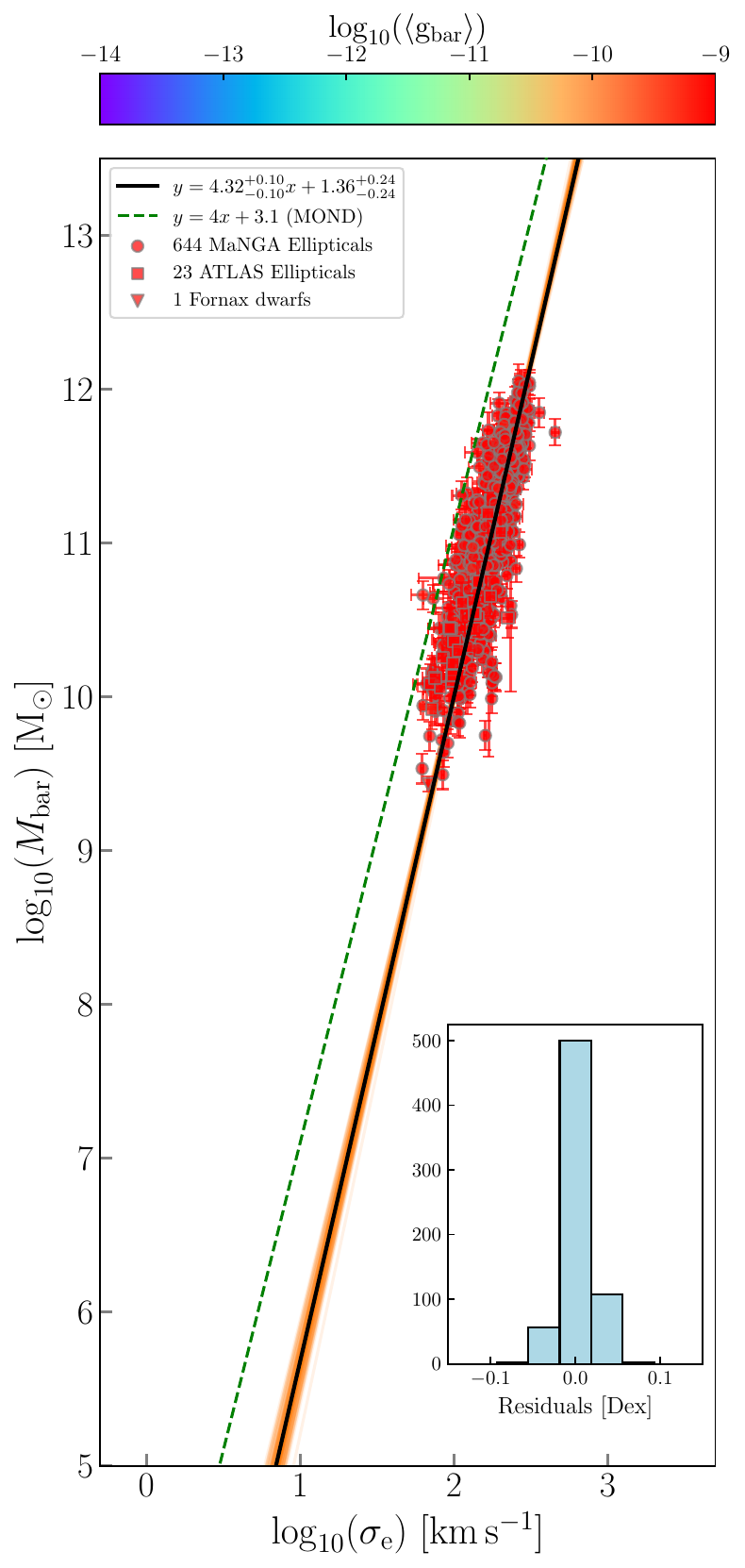}
\includegraphics[width=0.6\columnwidth]{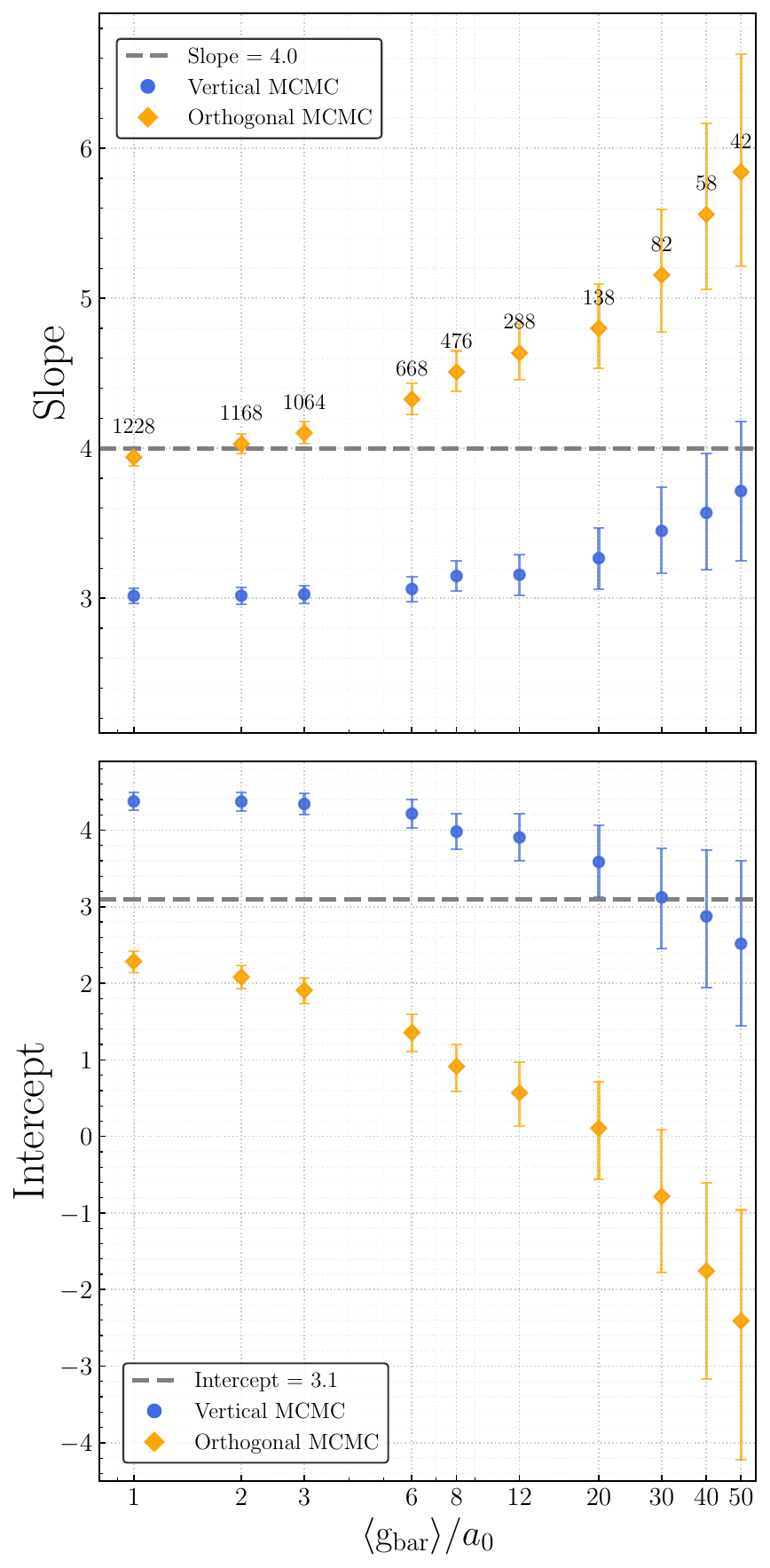}
\caption{\textit{Left panel:} The BFJR for the full sample.
\textit{Middle panel:} The BFJR for the high-acceleration subsample ($\gbarmed > 6\aZero$). In both panels, the black line shows the best-fitting relation from the orthogonal MCMC fit and the orange region denotes the $1\sigma$ credible interval. \textit{Right:} Variation of the fitted slope and intercept as a function of the acceleration threshold
$\gbarmed/\aZero$.}
\label{fig:BFJR_high}
\end{figure*}

The behavior of the BFJR under increasingly stringent acceleration thresholds is summarized in Figure~\ref{fig:BFJR_high}. The left panel reproduces the full sample, while the middle panel shows the high-acceleration subsample ($\gbarmed>6\aZero$) together with the best-fitting orthogonal MCMC relation, $y = 4.32x + 1.36$. The right panel compares orthogonal and vertical MCMC fits for subsamples selected above increasing thresholds of $\gbarmed/\aZero$, complementary to the presentation in the right panel of Figure~\ref{fig:BFJR}. The two fitting approaches yield systematically different results in this regime, indicating that the inferred BFJR parameters depend on the adopted fitting scheme. This is common for linear relations with steep slopes \citep[e.g.,][]{Lelli2019, Tian2021a}. 

\begin{figure*}[t]
\centering
\includegraphics[width=1.95\columnwidth]{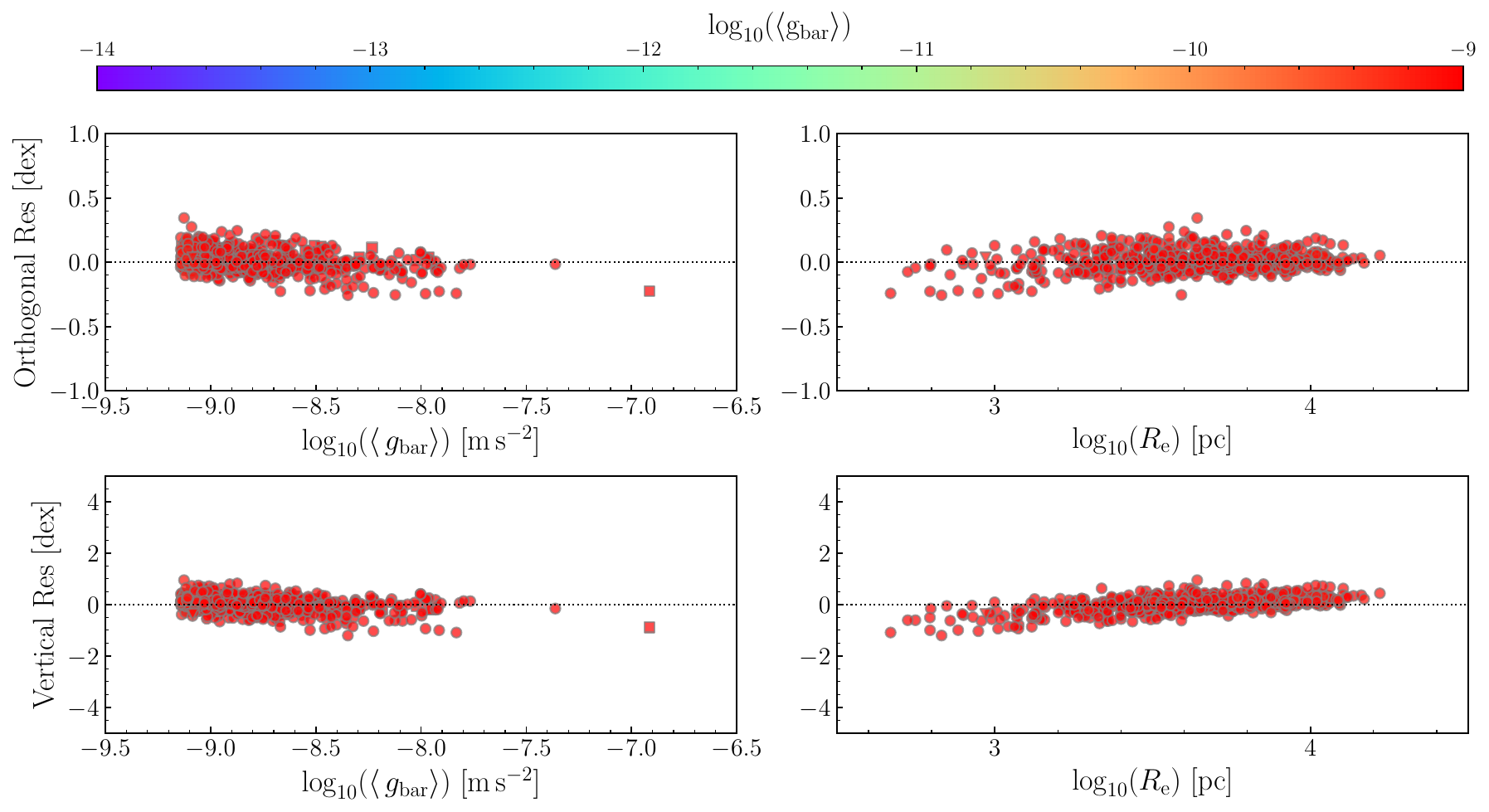}
\caption{BFJR residuals versus two different physical quantities for the high-acceleration sub-sample ($\gbarmed>6\aZero$). \textit{Top panels:} Orthogonal residuals versus the logarithm of the internal acceleration, $\log_{10}(\gbarmed)$ (left panel) and the logarithm of the effective radius, $\log_{10}(\Reff)$ (right panel).
\textit{Bottom panels:} Same as the top panels, but for the vertical residuals.}
\label{fig:Res_high}
\end{figure*}

To revisit the existence of the FP, Figure~\ref{fig:Res_high} presents both the
orthogonal and vertical residuals of the high-acceleration subsample as a function of internal acceleration and effective radius. It is evident by eye that the residuals correlate with $R_{\rm e}$ (as expected due to the FP) and more weakly with $\gbarmed$ (also expected because the internal baryonic acceleration depends on the baryonic surface density). Indeed, the Pearson's test gives a correlation coefficient $r \simeq 0.1$ with $p \simeq 2 \times 10^{-4}$ for the orthogonal residuals against $R_{\rm e}$, and  $r \simeq 0.58$ with $p \ll 10^{-5}$ for the vertical ones. This analysis confirms that an additional structural parameter can reduce the scatter around the BFJR of the high-acceleration regime, contrary to the case of the BFJR for the low-acceleration sample (see Fig.\,\ref{fig:Res}).

\section{Posterior probability distributions}\label{app:corner}
\begin{figure*}[!htb]
  \centering
  \includegraphics[width=0.95\columnwidth]{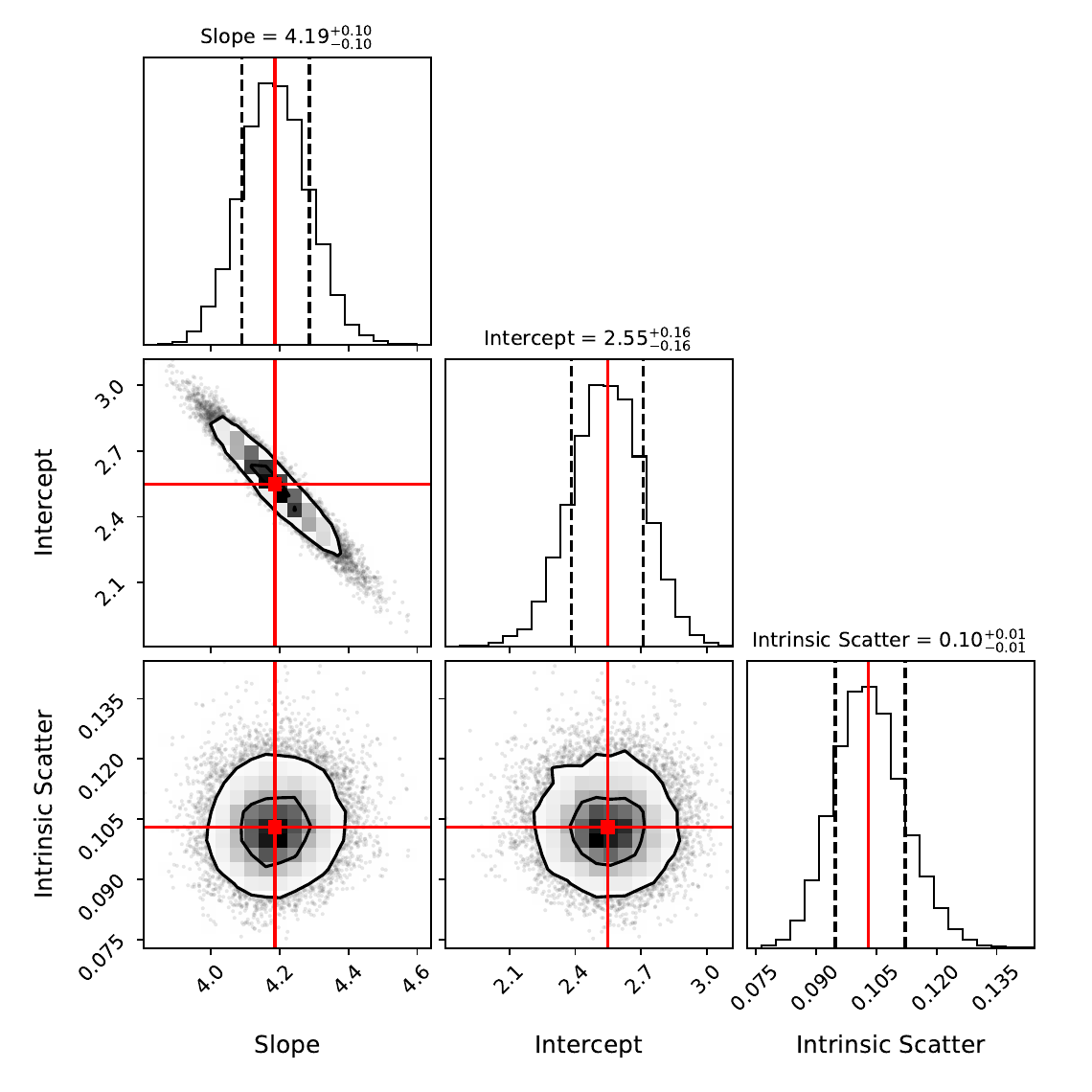}
  \includegraphics[width=0.95\columnwidth]{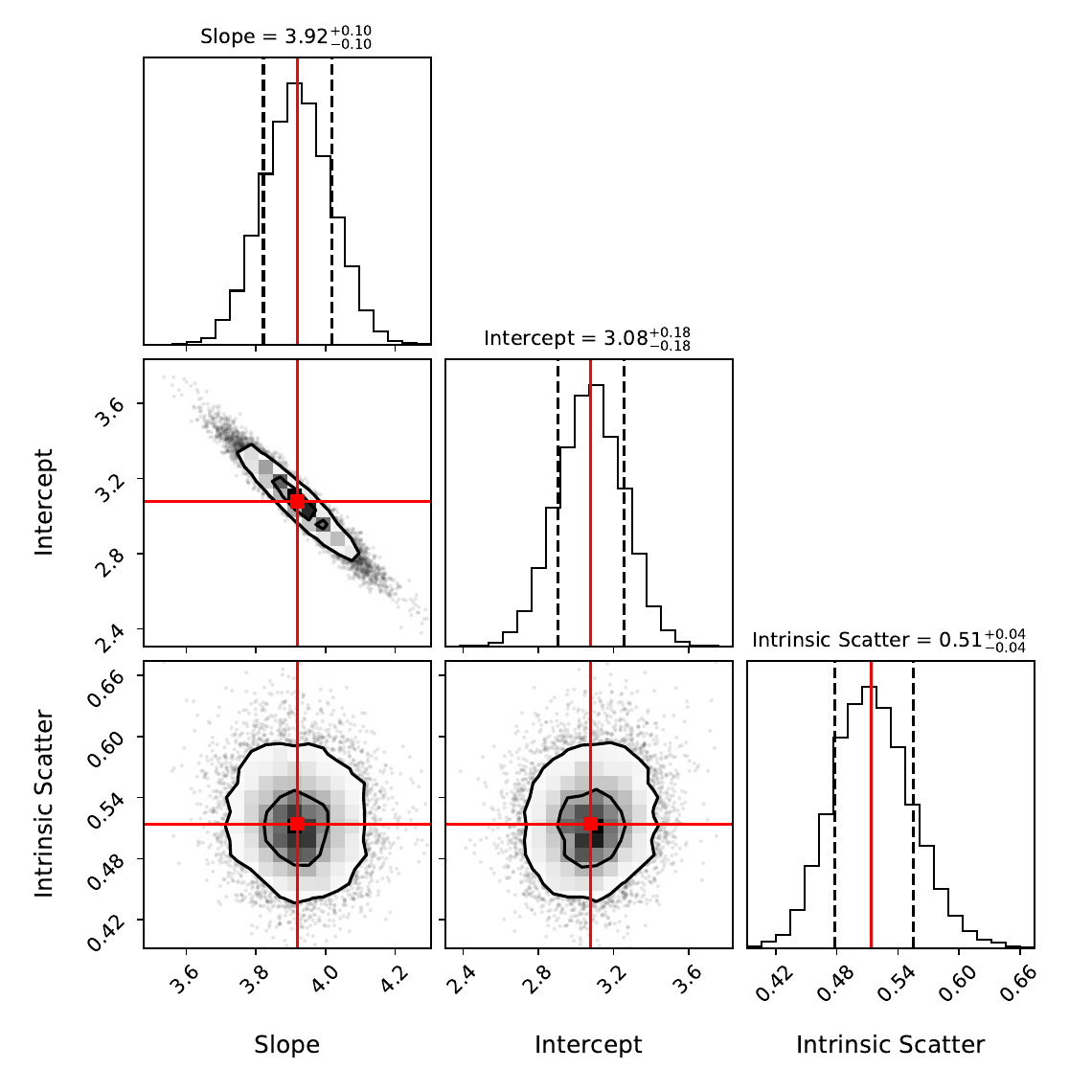}
  \caption{Posterior probability distributions of BFJR fit parameters selecting galaxies with $\gbar<0.6a_0$.
    \textit{Top:} Results from the orthogonal MCMC fit, showing the marginalized and joint posterior distributions for the slope, intercept, and intrinsic scatter of the BFJR. 
    \textit{Bottom:} Results from the vertical MCMC fit using the same data. 
    The best-fit values and $1\sigma$ uncertainties are indicated by red lines and annotations. Contours correspond to the $1\sigma$ and $2\sigma$ credible regions.
    The comparison demonstrates the impact of the fitting method on the derived BFJR parameters and their uncertainties.
}
  \label{fig:MCMC}
\end{figure*}

Figure~\ref{fig:MCMC} shows the ``corner plots'' for the orthogonal and vertical MCMC fits to the low-acceleration subsample ($\gbar<0.6\aZero$). The 1D posterior probability distributions are single-peaked and close to a Gaussian function, so the best-fit parameters and their uncertainties are well defined. As it always happens in linear fits, slope and intercept are somewhat degenerate.

\section{Residuals of the low-acceleration baryonic Faber-Jackson relation}\label{app:residuals}

\begin{figure*}[!htb]
\centering
\includegraphics[width=1.90\columnwidth]{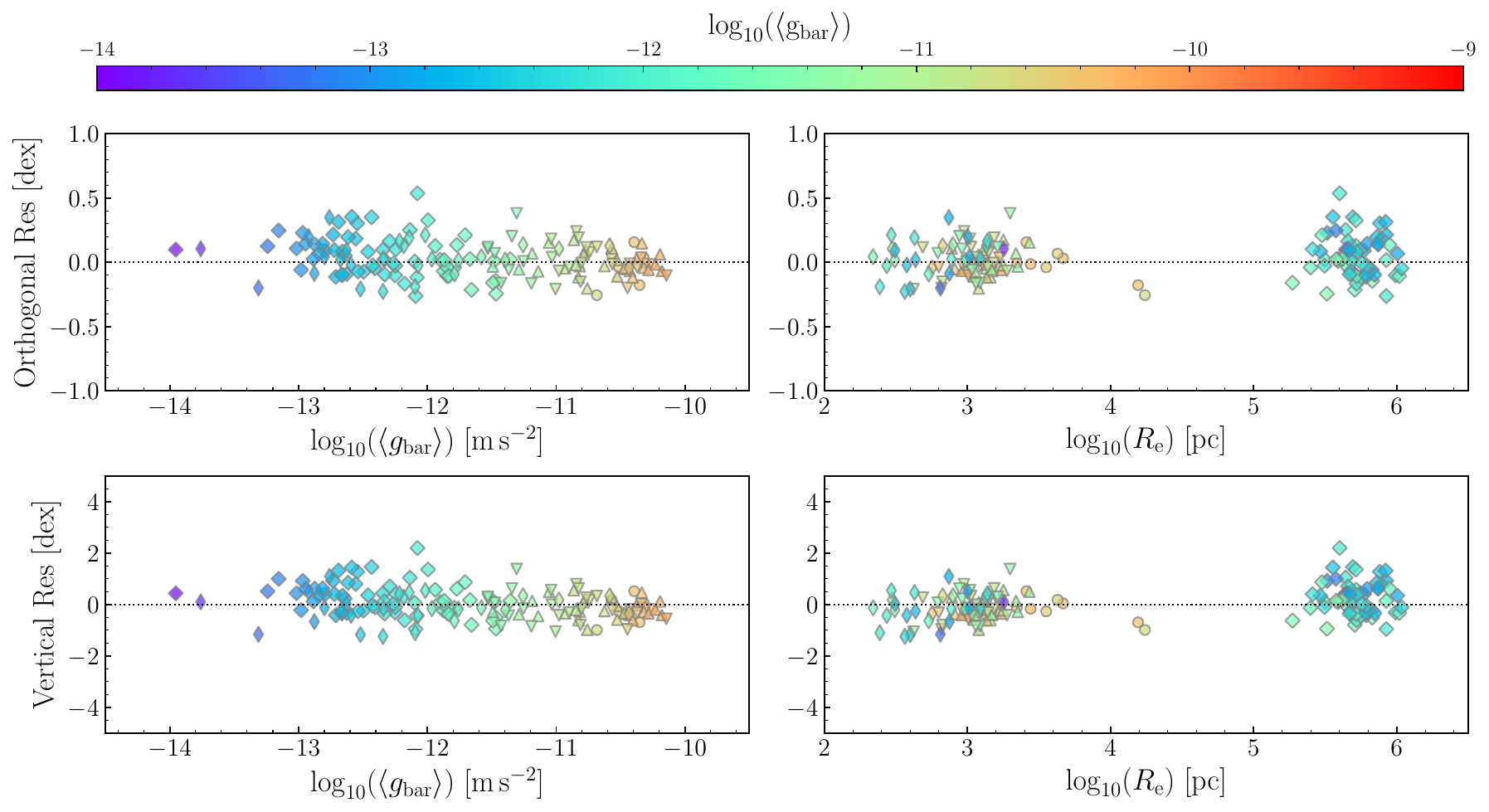}
\caption{BFJR residuals versus two different physical quantities for the low-acceleration subsample ($\gbarmed<0.6\aZero$).
\textit{Top panels:} Orthogonal residuals versus the logarithm of the mean internal acceleration, $\log_{10}{\gbarmed}$ (left panel) and the logarithm of the effective radius, $\log_{10}(\Reff)$ (right panel).
\textit{Bottom panels}: Same as the top panels, but for the vertical residuals.}
\label{fig:Res}
\end{figure*}

To check for potential secondary correlations, Fig.~\ref{fig:Res} shows the residuals of the low-acceleration BFJR against different properties of the systems. The orthogonal residuals show no systematic correlation with the characteristic size of the system (the effective radius of galaxies or the mean radius of galaxy groups). This indicates that the scatter around the BFRJ of low-acceleration systems cannot be decreased by adding a third structural variable, contrary to the case of the FP of high-acceleration systems (see Appendix \ref{app:High_acc}). Indeed, the Pearson's test gives a negligible correlation coefficient $r \simeq 0.1$ with $p\simeq0.14$. The same test gives a potentially significant correlation between vertical residuals and $R_{\rm e}$ ($r \simeq 0.3$ and $p\simeq5\times10^{-5}$) but this may be a shortcoming of the vertical fit, which is known to underestimate the best-fit slope of steep linear relations \citep{Lelli2019}.

Intriguingly, we find a statistically significant, albeit weak, anti-correlation with the internal acceleration $\gbarmed$ for both the orthogonal fit ($r \simeq -0.2$ with $p \simeq 5 \times 10^{-3}$) and for the vertical fit ($r \approx -0.3$ with $p \simeq 6 \times 10^{-4}$), which may possibly be driven by the external field effect in MOND, as we discussed in Sect.~\ref{sec:discussion}.

\section{Data availability}
Datasets are available in electronic form at the CDS via anonymous ftp to cdsarc.u-strasbg.fr (130.79.128.5) or via http://cdsweb.u-strasbg.fr/cgi-bin/qcat?J/A+A/

\FloatBarrier
\end{document}